\documentstyle[epsf,12pt,aaspp4,amssym,flushrt,tighten]{article}
\lefthead{Chau}
\righthead{Rotation And Magnetic Evolution Of Superconducting Strange Stars}
\slugcomment{IASSNS-HEP-96/16}
\begin{document}
\singlespace
\title{On The Rotation And Magnetic Field Evolution Of Superconducting Strange
 Stars}
\author{H. F. Chau\altaffilmark{1}}
\affil{School of Natural Sciences, Institute for Advanced Study, Olden Lane,
 Princeton, NJ 08540}
\altaffiltext{1}{e-mail: chau@sns.ias.edu}
\authoremail{chau@sns.ias.edu}
\begin{abstract}
 Is pulsar make up of strange matter? The magnetic field decay of a pulsar may
 be able to give us an answer. Since Cooper pairing of quarks occurs inside a
 sufficiently cold strange star, the strange stellar core is superconducting.
 In order to compensate the effect of rotation, different superconducting
 species inside a rotating strange star try to set up different values of
 London fields. Thus, we have a frustrated system. Using Ginzburg-Landau
 formalism, I solved the problem of rotating a superconducting strange star:
 Instead of setting up a global London field, vortex bundles carrying localized
 magnetic fields are formed. Moreover, the number density of vortex bundles is
 directly proportional to the angular speed of the star. Since it is
 energetically favorable for the vortex bundles to pin to magnetic flux tubes,
 the rotational dynamics and magnetic evolution of a strange star are coupled
 together, leading to the magnetic flux expulsion as the star slows down. I
 investigate this effect numerically and find that the characteristic field
 decay time is much less than 20~Myr in all reasonable parameter region. On the
 other hand, the characteristic magnetic field decay time for pulsars is $\geq
 20$~Myr. Thus, my finding cast doubt on the hypothesis that pulsars are
 strange stars.
\end{abstract}
\section{Introduction \label{S:Intro}}
 What is the most stable form of baryonic matter at high density? This question
 is of both observational and theoretical interests. With the discovery of
 pulsars and their identification with the neutron stars in the late 60s, many
 people thought that neutron star matter is the most stable form of cold
 condensed matter at high density. This believe was later challenged by Witten.
 He proposed that at sufficiently high density, deconfinement of nucleons
 occurs. Moreover, some of the d and u quarks on of the Fermi surface is
 converted to strange quarks via weak interaction. Thus, the most stable form
 of matter at ultra-high density is a degenerate mixture of u, d, and s quarks
 together with a small amount of electrons so as to maintain overall charge
 neutrality (\cite{Witten}). This kind of material state is called strange
 matter. His idea was examined in detail later by Farhi and Jaffe (1984).
\par
 The possibility of self gravitating strange matter, namely, a strange star,
 has also been investigated. The equation of state (EOS) calculations of
 strange star suggests that they are indeed stable objects in certain parameter
 range (\cite{Alcock86}; \cite{Haensel86}; \cite{Benvenuto89}; see also
 \cite{Proc91}; and \cite{Strange_Pulsar_EOS}). The mass and radius of a stable
 strange star are found to be similar to those of a neutron star. Consequently,
 some authors suggested that pulsars are in fact strange stars (see for example
 \cite{Strange_Pulsar_EOS}). In principle, the question of whether pulsars are
 neutron or strange stars can be answered by comparing the core density of a
 neutron star to the strange matter transition density. Unfortunately, all
 the nuclear physics calculations to date do not yield a definitive answer.
\par
 There are a number of difficulties in explaining some of the observational
 properties of pulsars using the strange star model. The first, and perhaps
 the most serious difficulty, is the possibility of a strange star glitch.
 Early EOS calculations suggested that strange stars have very thin nuclear
 matter crust. The ratio of inertial moment of the nuclear matter crust to that
 of the whole strange star is of the order of $10^{-5}$. Moreover, the density
 of the nuclear matter crust is not high enough for neutron drip
 (\cite{Alcock86}; \cite{Alcock91}). In contrast, pulsar glitch observations
 tell us that the ratio of inertial moment of pinned neutron superfluid crust
 to that of the whole star is about $10^{-2}$ (\cite{Glitch_Comment};
 \cite{Link}; \cite{VCM}). Nevertheless, by taking into account the existence
 of strange matter bound states, Benvenuto et al. (1991ab) showed that strange
 star may support a ``strange matter crust'' thick enough to account for the
 observed pulsar glitches. Besides, a more recent calculation suggested that a
 strange star may be able to retain a reasonable size of nuclear matter crust
 by accretion as well (\cite{Benvenuto94}).
\par
 The second difficulty is the possibility of type~I X-ray bursts on a strange
 star surface. It is commonly believed that type~I X-ray burst involves sudden
 thermo-nuclear runaway at the surface of an accreting neutron star (see for
 example \cite{X_Burst}). The same bursting behavior cannot occur on a bare
 strange star surface because nuclear fuel will dissociate into constituent
 quarks immediately (\cite{Jones86}). The possibility of X-ray burst,therefore,
 requires the existence of a nuclear matter crust over a strange star. EOS
 calculations showed that the nuclear matter crust is separated from the
 interior strange matter by a electrostatic gap of a few fermi thick, thereby
 preventing the nuclear matter from converting into strange matter
 (\cite{Alcock86}; \cite{Curst_SP}). A strange pulsar magnetosphere can also be
 formed, giving rise to the observed radio pulsation (\cite{Alcock86};
 \cite{Strange_Pulsar_EOS}).
\par
 The final difficulty is the magnetic field decay time. Assuming a {\em normal}
 strange matter core, Jones (1988) pointed that magnetic field inside a strange
 star will not decay in Hubble's time. Moreover, there is no obvious mechanism
 for the star to retain a small residual field after say $10^9$ years.
 Therefore, it is inconsistent with both the pulsar magnetic field decay
 hypothesis, and the spun-up formation scenario of milli-second pulsars
 (\cite{Jones88}). Nevertheless, the statistical evidence for pulsar magnetic
 field decay over a period of some $10^9$~yr is still inconclusive. Also,
 milli-second pulsars may rather form by accretion induced collapse of white
 dwarfs (see for example \cite{Field_Decay} and \cite{Decay4} for discussions
 on neutron star magnetic field decay). Therefore, the objection of Jones may
 not be completely well founded. Moreover, as I shall discuss in \S\ref{S:SM},
 the core of a strange star is likely to be superconducting. In this case, the
 magnetic field decay time estimated by Jones (1988) has to be modified. The
 question of magnetic field decay have been brought up in the review paper by
 Bailin and Love (1984). They briefly mentioned that the low electron density
 in the strange stellar core may lead to rapid flux expulsion. However, they
 did not provide a detail calculation. A major objective of this paper,
 therefore, is to perform such a calculation, taking into account the coupling
 between magnetic evolution and rotational dynamics of the star in the presence
 of collective effects such as clumping of flux tubes.
\par
 At this moment, one has no doubt that the standard neutron star model for
 pulsars is well tested by numerous observational data. However, we still have
 to keep a close look at the alternative hypothesis, namely, the strange pulsar
 model (\cite{Strange_Pulsar_EOS}). In particular, we like to explore various
 physical properties of a neutron star and a strange star, and to see if there
 are further ways to test if which of the two models is a better candidate in
 explaining pulsar observations.
\par
 As I shall discuss in \S\ref{S:SM}, the core of a strange star is likely to be
 superconducting (\cite[1984]{Bailin82}). However, this cast a problem on the
 rotation of a strange star. In order to rotate an object with one
 superconducting species, a uniform magnetic field
\begin{equation}
 {\bf B} = \frac{2m^* c}{q^*}\,{\bf\Omega} \mbox{~,} \label{E:London}
\end{equation}
 called the London field, is set up in the object (\cite{London_Theo}), where
 $m^*$ and $q^*$ are the effective mass and charge of the Cooper pair, and
 $\Omega$ is the rotational angular velocity of the object. But for a strange
 star, u, d, and s quarks are all superconducting. These superconducting
 species require different values of $B$ to set up a rotation. Thus, we have a
 {\em frustrated} system. The situation is further complicated by the fact that
 all three superconducting quark flavors interact strongly with each other. In
 \S\ref{S:Rotate}, I tackle this rotation problem using Ginzburg-Landau
 formalism. Instead of setting up a uniform London field, I find that vortices
 and localized magnetic fields are created when the superconducting quarks
 rotate together. Then in \S\ref{S:Ap}, I point out that the vortices will
 inter-pin with the magnetic flux tubes, which alter the magnetic field
 evolution and rotation of a strange star dramatically. In particular, it
 suggests that the (superconducting) strange pulsar hypothesis is inconsistent
 with the observed magnetic field decay time in pulsars. Finally, a conclusion
 is drawn in \S\ref{S:Conclude}.
\section{Strange Matter Superconductivity \label{S:SM}}
 Using a relativistic treatment of the BCS theory, Bailin and Love (1982, 1984)
 suggested that strange matter turns superconducting at low temperatures. Using
 perturbative QCD at the one gluon exchange level, they showed that the pairing
 of quarks is most likely to occur in both the ud-du and the ss channels. The
 ss pairing is expected to have a gap matrix transforming as a color
 $\overline{\underline{3}}$ and having $J^P = 1^+$ (and hence the pairing is in
 p-wave). On the other hand, pairing in the u-d system is expected to occur in
 the isoscalar channel, with a gap matrix transforming as a color $\overline{
 \underline{3}}$ and having $J^P = 0^+$ (and hence the pairing is in s-wave)
 (\cite{Bailin84}). The superconducting transition temperature is about
 $400$~keV. Incidentally, the transition temperatures for neutron superfluid
 and proton superconductor in a neutron star core are of the same order of
 magnitude. Thus, about 1000~yr after its supernova birth, the interior of a
 typical strange star is already cold enough for quark superconductivity
 (\cite{SS_Cooling}). However, one should notice that the above estimation
 depends quite sensitively on the quark-gluon coupling, and may also change if
 one goes beyond the one gluon exchange calculations. Thus, the existence of a
 superconducting core in a strange star is not completely conclusive even
 though it is very likely. Nonetheless, I assume the existence of quark
 superconductivity in a strange star core in this paper.
\par
 The quark superconductor is likely to be marginally type~I with a zero
 temperature critical field $B_c$ about $10^{16}$~G (\cite{Bailin84}).
 Incidentally, the lower critical field of proton superconductor at the core of
 a neutron star is also of this order. The typical magnetic field of a
 canonical pulsar is about $10^{12}$~G, so naively one would expect complete
 flux expulsion from the superconducting strange matter core. However, the huge
 electrical conductivity of the normal state opposes the motion of the flux,
 thereby, leading to a long flux expulsion time. In the case of a neutron star
 core, the flux expulsion time is $\sim 10^8$~years (\cite{Baym69}). This time
 is shorter for strange stars because of their low electron density and strong
 quark-quark scattering amplitude. Bailin and Love (1982) predicted that it may
 be as little as $10^4$~yr. However, their estimate did not take into account
 the possibility of flux clumping, which may greatly increase the flux
 expulsion time. Moreover, if the flux expulsion rate is so fast, then a
 Meissner state is formed in an old pulsar. As we shall point out in
 \S\ref{S:Stru}, the thin crustal nuclear matter is not strong enough to
 support the magnetic stress and tension of the expelled field. Therefore, we
 believe that during the life of a canonical pulsar, a meta-stable state with
 magnetic field penetrating through the superconducting strange star core has
 to be formed. We shall return to this point later in \S\ref{S:Ap}.
\par
 Since the quark superfluid also couples to the color vector gluons, color
 superconductivity may also be observed. Because of color confinement outside
 the stellar core, we do not have the color analogue of the ambient $10^{12}$~G
 magnetic field (\cite{Bailin84}). Thus, the observational consequences of
 color superconductivity remains unclear.
\section{Rotating Strange Star --- Ginzburg-Landau Approach \label{S:Rotate}}
 As I have discussed in \S\ref{S:Intro}, London magnetic field is set up when a
 single species of superconducting sample is rotated (\cite{London_Theo};
 \cite{Leggett}). In fact, the presence of London field has been observed in
 terrestrial superconductors (\cite{London_Expt}). In \S\ref{S:Rotate_1}, we
 re-derive the expression in Eq.~(\ref{E:London}) using Ginzburg-Landau
 formalism as a warm up to the multi-superconducting-component situation like
 that in a strange star. In the derivation, I find that there is a critical
 angular velocity above which the uniform London field state is not the most
 energetically stable configuration. This critical angular velocity, as far as
 I know, has not been investigated in the literature. Unfortunately, such a
 critical angular velocity is too high to be attained experimentally.
\subsection{The Case When There Is Only One Superconducting Species
 \label{S:Rotate_1}}
 We assume that the Cooper pairs in the sample are in the s-state. Thus, the
 order parameter for the only superconducting species $\Psi$ is complex. In
 case the Cooper pairs are in higher angular momentum state, the order
 parameter will be a complex matrix (\cite{Vollhardt}). We, however, shall only
 stick to the simple s-wave pairing derivation. Similar results also apply to
 higher angular momentum pairing states although the derivation becomes rather
 complicated. For a sufficiently small angular velocity $\Omega$ and sample
 size, it is reasonable that the speed of the superfluid is not high at any
 point and hence a non-relativistic treatment will suffice. In the co-rotating
 frame, the Ginzburg-Landau energy functional reads as $F = \int f\,dV$ where
\begin{equation}
 f = - a |\Psi |^2 + \frac{b}{2} |\Psi |^4 + \frac{1}{2m^*} \left| \left(
 \frac{\hbar{\bf\nabla}}{i} + \frac{q^*}{c} {\bf A} - m^* {\bf\Omega} \times
 {\bf r} \right) \Psi \right|^2 + \frac{1}{8\pi} \left| {\bf\nabla}\times
 {\bf A} \right|^2 \mbox{~.} \label{E:GL_1}
\end{equation}
 Here, $m^*$ and $q^*$ are the {\em effective} mass and charge of the Cooper
 pairs. Moreover, $a, b > 0$ so that the superconducting state is prefered over
 the normal state. In general, $a$, and $b$ are temperature dependent. However,
 a few hundred years after its supernova birth the cooling timescale for a
 strange star is much longer than its spin-down timescale. Thus, as far as
 rotational dynamics of the star is concerned, we may assume that $a$ and $b$
 are constants.
\par
 We consider the case when there is no external magnetic field. The presence of
 an external magnetic field may simply create extra fluxoids in the sample,
 which is not very interesting. There are two ways to minimize the kinetic
 energy term in Eq.~(\ref{E:GL_1}). First, we can set up a magnetic field so
 that the ${\bf\Omega}\times{\bf r}$ contribution is cancelled by ${\bf A}$.
 Alternatively, we can create normal cores (i.e. line defects in the form of
 vortices similar to that of a rotating superfluid). However, there are prices
 to pay for both cases: setting up magnetic field requires magnetic energy;
 setting up vortex needs both a kinetic energy near the vortex and a nucleation
 energy for creation of the vortex core. What really happens in the sample, in
 general, is that a uniform magnetic field together with an array of vortices
 will be formed when it is rotated. Suppose a uniform magnetic field ${\bf B}$
 (parallel to ${\bf\Omega}$) is set up in the star, then ${\bf A} = {\bf B}
 \times{\bf r} / 2$. Then in order to minimize the kinetic energy term in
 Eq.~(\ref{E:GL_1}), vortices (or anti-vortices) have to be formed just like a
 rotating superfluid (see for example \cite{Sauls_Review}). Unlike fluxoids,
 vortices carries circulation but not magnetic field. The number of vortex (or
 anti-vortex) per unit area required is given by
\begin{equation}
 n(B) = \frac{2}{\kappa} \left| \Omega - \frac{q^*}{2m^* c} B \right| \mbox{~,}
 \label{E:density}
\end{equation}
 where $\kappa = h / m^*$ is the circulation quantum. Obviously, vortices forms
 an Abrikosov lattice in the absence of spatial inhomogeneity. The
 Ginzburg-Landau free energy per unit volume of the system is, therefore, given
 by (\cite{Sauls_Review}; \cite{Vollhardt})
\begin{equation}
 \frac{F}{V} \approx \frac{B^2}{8\pi} + \left\{ \frac{\kappa^2 \rho_s}{4\pi}
 \left[ \ln \left( \frac{R_c}{\xi} \right) - \frac{3}{4} \right] + \frac{1}{2}
 \,\pi \xi^2 N(0) \Delta^2 \right\} \,n(B) \mbox{~,} \label{E:F_1}
\end{equation}
 where $\rho_s = m \Psi^* \Psi$ is the density of superconducting species,
 $\xi$ is the coherence length of the vortex, $R_c$ is the upper cutoff length
 which is of the order of the inter-vortex spacing, $\Delta$ is the
 superconducting gap energy, and $N(0)$ is the density of state for one spin
 projection on the Fermi surface which is given by
\begin{equation}
 N(0) = \left[ k^2 \frac{dk}{dE_k} \right]_{k=k_F} \approx \frac{E_F^3}{2\pi^2
 \hbar^3 c^3} \label{E:N0_def}
\end{equation}
 in the extreme relativistic limit.
\par
 Physically, the first term in Eq.~(\ref{E:F_1}) is the magnetic energy, and
 the second term is the kinetic and nucleation energies of the vortices. By
 minimizing Eq.~(\ref{E:F_1}), we find that the magnetic field generated in the
 star is
\begin{equation}
 B = \left\{ \begin{array}{cl} \displaystyle \frac{2m^* c}{q^*}\,\Omega &
 \hspace{0.25in} \mbox{if~} \Omega < \Omega_c \vspace{0.2in} \\ \displaystyle
 \frac{q^*}{m^* c\kappa} \left\{ \kappa^2 \rho_s \left[ \ln \left(
 \frac{R_c}{\xi} \right) - \frac{3}{4} \right] + 2\pi^2 \xi^2 N(0) \Delta^2
 \right\} & \hspace{0.25in} \mbox{otherwise} \end{array} \right. \mbox{~,}
 \label{E:B_1}
\end{equation}
 where we have neglected the dependence of $\ln R_c$ on $B$ because it is
 rather weak. The critical angular velocity $\Omega_c$ above which vortices
 begin to form and the translational symmetry of the system is spontaneously
 broken can be determined from Eqs.~(\ref{E:London}) and~(\ref{E:B_1}). And it
 is given by
\begin{equation}
 \Omega_c = {\displaystyle \frac{{q^*}^2 \rho_s h}{2{m^*}^3 c^2}} \mbox{~.}
 \label{E:Omega_c_1}
\end{equation}
\par \indent
 Since magnetic energy scales quadratically with magnetic field while the
 energy required to form an array of vortices scales linearly with it,
 formation of vortices is favored at high rotational rates. For a typical
 terrestrial superconductor (high or low $T_c$), the effective charge and mass
 of the Cooper pairs are given by $q^* = -2e$ and $m^* = 2m_e$ respectively. By
 putting $\rho_s \sim 10^{-6}$~g\,cm$^{-3}$, we obtain $\Omega_c \sim 6\times
 10^8$~rad\,s$^{-1}$. Similarly, for neutron star matter, $q^* = 2e$, $m^* = 2
 m_p$, and $\rho_s \sim 2\times 10^{12}$~g\,cm$^{-3}$, $\Omega_c \sim 2\times
 10^{17}$~rad\,s$^{-1}$. $\Omega_c$ for strange star matter is similar to that
 of a neutron star. These angular velocities, unfortunately, are so high that
 the sample or the star will break apart well before the formation of vortices.
 In other words, we can neglect the magnetic energy contribution in the
 Ginzburg-Landau free energy all the time.
\subsection{The Case When There Are Multiple Non-interacting Superconducting
 Species \label{S:Rotate_m}}
\begin{figure}
 \plotone{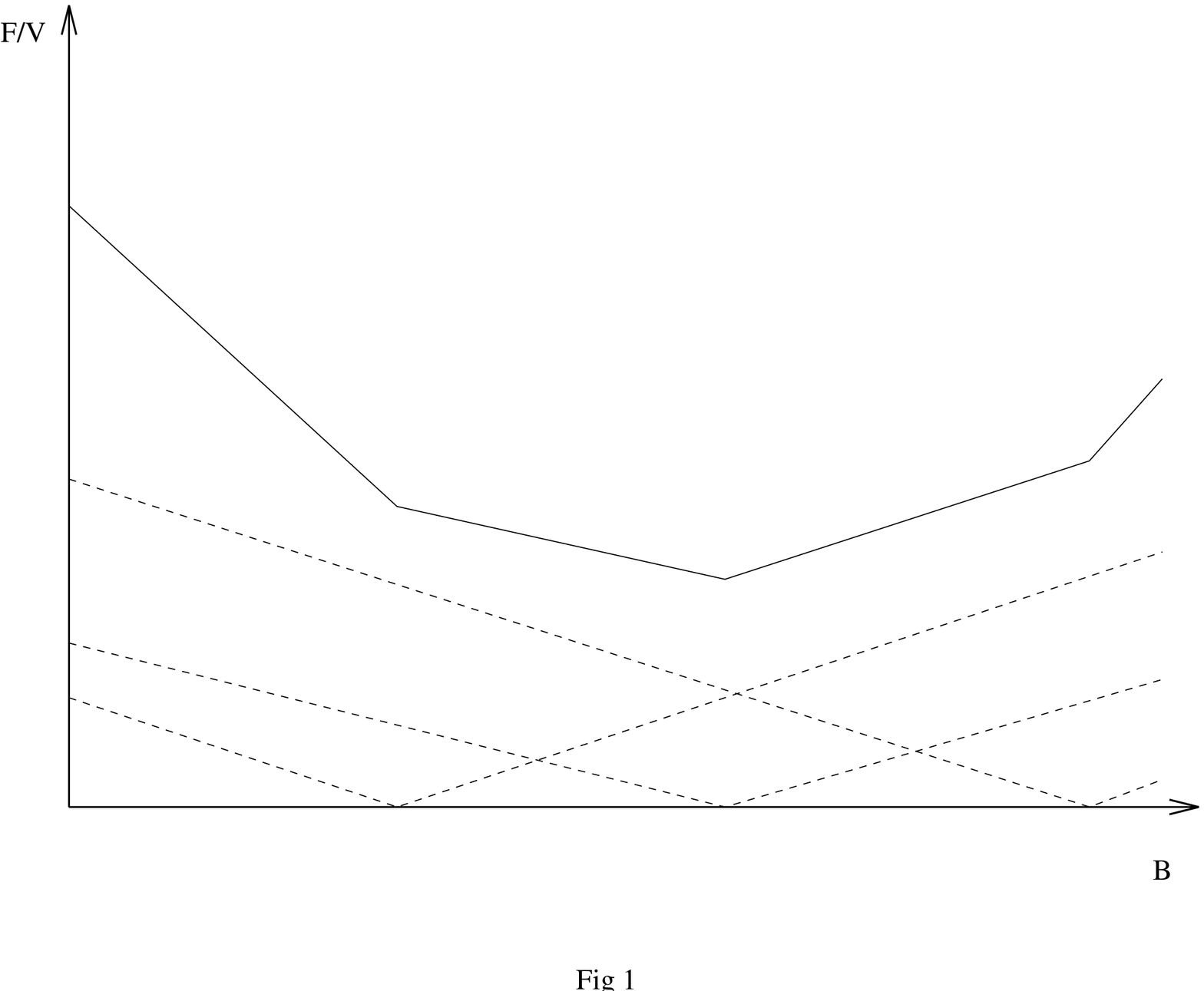}
 \caption{Schematic plot of $F / V$ as a function of $B$ for a fixed $\Omega$
  in a multi-superconducting non-interacting species sample. Dash lines
  represent the contributions from individual superconducting species, and the
  solid line is their combined free energy per unit volume. \label{F:F_V}}
\end{figure}

 Now we examine the case when there are more than one superconducting species.
 We consider the simple case when they are non-interacting. Thus, the only
 coupling between them is that they response to the {\em same} vector potential
 ${\bf A}$. If we label each superconducting species by $j$, the
 Ginzburg-Landau free energy density is
\begin{equation}
 f = \sum_j \left\{ -a_j |\Psi_j |^2 + \frac{b_j}{2} |\Psi_j |^4 +
 \frac{1}{2m_j^*} \left| \left( \frac{\hbar {\bf\nabla}}{i} + \frac{q_j^*}{c}
 {\bf A} - m_j^* {\bf\Omega} \times {\bf r} \right) \Psi_j \right|^2 \right\} +
 \frac{1}{8\pi} \left| {\bf\nabla}\times {\bf A} \right|^2 \mbox{~.}
 \label{E:GL_m}
\end{equation}
 Once again, we expect each superconducting species to form vortices in
 addition to the generation of a common magnetic field. (Compare with Mendell
 and Lindblom (1991) for a similar study using hydrodynamics.) As we have shown
 in \S\ref{S:Rotate_1}, we can neglect the magnetic energy contribution. As a
 result, for a uniform magnetic field $B$, the Ginzburg-Landau free energy per
 unit volume becomes
\begin{equation}
 \frac{F}{V} \approx \sum_j \left\{ \frac{\hbar \rho_{sj}}{m_j^*} \,\left[ \ln
 \left( \frac{R_{cj}}{\xi_j} \right) - \frac{3}{4} \right] + \frac{m_j^*
 \xi_j^2}{2 \hbar} N_j(0) \Delta_j^2 \right\} \,\left| \Omega - \frac{q_j^*}{2
 m_j^* c} B \right| \mbox{~.} \label{E:F_m}
\end{equation}
 As shown in Fig.~\ref{F:F_V}, for a fix $\Omega$, the free energy density
 contribution of each superconducting species assumes its minimum value when
 $B$ equals its London magnetic field. Neglecting the weak dependence of $\ln
 R_c$ on $B$, $F / V$ is a continuous and piecewise linear function of $B$ with
 ``vertices'' locate at the points where $B$ equals the London magnetic field
 of any one of its superconducting species. Thus, the actual magnetic field $B$
 set up in the sample always equals to a London magnetic field of a particular
 superconducting species. In other words, at least one superconducting species
 rotates {\em without} creating vortices. While other species with a different
 London field have to create vortices (or anti-vortices) in order to rotate
 with the same angular velocity. This finding is somewhat unexpected because
 one would naively think that the value of the magnetic field set up in the
 sample would be a compromise between different London fields and all
 superconducting species would form vortices. As the superconducting species do
 not interact, vortices from each species will form Abrikosov lattices on their
 own (in the absence of spatial inhomogeneity).
\par
 What we can predict if the quarks in strange stars were non-interacting? Since
 the masses (let it be bare or consitutent) of d and u are about the same,
 which are both much less than that of s, Eq.~(\ref{E:London}) tells us that
 their corresponding London fields satisfy
\begin{equation}
 B_s < 0 < B_{ud} \label{E:Order_m}
\end{equation}
 when $\Omega > 0$. (An exactly reverse relationship is true when $\Omega <
 0$.) Since $\rho_u / m_u^* \approx \rho_d / m_d^* \gg \rho_s / m_s^*$
 (\cite{Alcock91}), Eq.~(\ref{E:GL_m}) tells us that $g_{ud} \gg g_s$ where
\begin{equation}
 g_j = \frac{\rho_j |q_j^*|}{2 {m_j^*}^2 c} \hspace{0.25in} \mbox{for~} j =
 \mbox{ud,s.} \label{E:g_j}
\end{equation}
 Combining Eqs.~(\ref{E:F_m}),~(\ref{E:Order_m}) and~(\ref{E:g_j}), the slope
 $M$ of our piecewise linear free energy density curve $F / V$ is given by
\begin{equation}
 M = \left\{ \begin{array}{rccl} g_s + g_{ud} & > & 0 & \hspace{0.1in}
 \mbox{for~} B > B_{ud} \\ g_s - g_{ud} & < & 0 & \hspace{0.1in} \mbox{for~}
 B_s < B < B_{ud} \\ -g_s - g_{ud} & < & 0 & \hspace{0.1in} \mbox{for~} B < B_s
 \end{array} \right. \mbox{~,} \label{E:Slope_m}
\end{equation}
 where we have neglected the small nucleation energy contribution in the above
 estimate.
\par
 Consequently, $F / V$ finds its minimum when $B = B_{ud}$. (Readers can verify
 that this conclusion is independent of the sign of $\Omega$.) So if different
 flavor quarks were non-interacting, only s quarks would form vortices upon
 rotation.
\subsection{The Case When There Are Multiple Interacting Superconducting
 Species \label{S:Rotate_mi}}
 Finally, we consider the (realistic) case when the quarks are interacting. The
 form of the free energy density must be invariant under global phase changes
 in any of the order parameters. If we consider terms up to second order in
 gradients and quartic in order parameters, the most general form of $f$ is
 given by (compare with Alpar et al. 1984b where they have omitted the quartic
 self-interaction terms)
\begin{eqnarray}
 f & = & \frac{1}{8\pi} |{\bf\nabla}\times {\bf A}|^2 + \sum_j \left\{ -a_j |
 \Psi_j |^2 + \frac{1}{2m_j^*} \left| {\cal P}_j \Psi_j \right|^2 \right\} +
 \frac{1}{2} \sum_{j,k} \left\{ b_{jk} |\Psi_j |^2 |\Psi_k |^2 +
 \right. \nonumber \\ & & \hspace{0.1in} \mu_{jk} ( {\cal P}_j \Psi_j^{*} )
 \cdot ( {\cal P}_k \Psi_k ) \Psi_k^{*} \Psi_j + \nu_{jk} ( {\cal P}_j
 \Psi_j^{*} ) \cdot ( {\cal P}_k \Psi_k^{*} ) \Psi_j \Psi_k + \nonumber \\ & &
 \hspace{0.1in} \left. \nu_{jk}^{*} ( {\cal P}_j \Psi_j ) \cdot ( {\cal P}_k
 \Psi_k ) \Psi_j^{*} \Psi_k^{*} \right\} \mbox{~,} \label{E:GL_mi}
\end{eqnarray}
 where $b$ is a real symmetric matrix, $\mu$ is an hermitian matrix, and
\begin{equation}
 {\cal P}_j = \frac{\hbar {\bf\nabla}}{i} + \frac{q_j^*}{c} {\bf A} - m_j^*
 {\bf\Omega} \times {\bf r} \label{E:Def_P}
\end{equation}
 is the covariant momentum operator in the co-rotating frame for the $j$-th
 species. Since the free energy has to be bounded from below, we further
 require all the eigenvalues of the symmetric matrix $b$ to be positive.
\par
 To explicitly show the drag effect of one superconducting species on the
 other, we write the order parameters $\Psi_j$ as $|\Psi_j | \exp (i\varphi_j
 )$, where $\varphi_j$ are the phases of the order parameters. Then, the
 velocity of the superconducting species $j$ in the rotating frame is given by
\begin{equation}
 {\bf v}_j = \frac{1}{m_j^* \Psi_j} {\cal P}_j \Psi_j \approx
 \frac{\hbar}{m_j^*} {\bf\nabla} \varphi_j + \frac{q_j^*}{m_j^* c} {\bf A} -
 {\bf\Omega}\times {\bf r} \mbox{~,} \label{E:Def_v}
\end{equation}
 where we have neglected the spatial variation of $|\Psi_j |$. From
 Eqs.~(\ref{E:GL_mi}) and~(\ref{E:Def_v}), we have
\begin{equation}
 f = \frac{1}{8\pi} |{\bf\nabla}\times {\bf A}|^2 - \sum_j a_j |\Psi_j |^2 +
 \frac{1}{2} \sum_{j,k} \left[ b_{jk} |\Psi_j |^2 |\Psi_k |^2 + \rho_{jk}
 {\bf v}_j \cdot {\bf v}_k \right] \mbox{~,} \label{E:f_drag}
\end{equation}
 where
\begin{mathletters}
\begin{equation}
 \rho_{jj} = m_j^* |\Psi_j |^2 + \mu_{jj} {m_j^*}^2 |\Psi_j |^4 + 2 \nu_{jj}
 {m_j^*}^2 |\Psi_j |^4 \label{E:rho_jj}
\end{equation}
and
\begin{equation}
 \rho_{jk} = \left( \mu_{jk} + 2 \nu_{jk} \right) m_j^* m_k^* |\Psi_j |^2
 |\Psi_k |^2 \label{E:rho_jk}
\end{equation}
\end{mathletters}
\noindent
 for $j \neq k$. Thus, once a superconducting species moves,
 Eq.~(\ref{E:f_drag}) tells us that in general it is energetically favorable
 for the other superconducting species to move along with it as well. A similar
 conclusion can be reached by using three-velocity hydrodynamics
 (\cite{Khalatnikov}; \cite{Andreev76}; \cite{Vardanyan81}).
\par 
 At the strange matter density, which is about $5\times 10^{14}$~gm\,cm$^{-3}$,
 the ``fine structure constant'' for strong force $\alpha_s$ is about 0.5 --
 0.6 (\cite{Strange_Pulsar_EOS}) indicating that QCD is the dominant
 interaction between the quark Cooper pairs. The values of the off-diagonal
 terms of the matrix $\rho$, which measure the strength of the drag, can be
 calculated (in principle) from the QCD interaction Hamiltonian. Although the
 large value of $\alpha_s$ prevents us from using perturbation theory, we
 believe that the effective coupling constants $\mu_{ij}$ and $\nu_{ij}$ in
 Eq.~(\ref{E:GL_mi}) are at least of order of $\alpha_s$. Thus, the drag force
 between different quark flavors are so strong that up to first order
 approximation, all three quark flavors move at the same velocity except
 possibly near the superfluid cores. The co-moving approximation greatly
 simplifies our effort to find the minimum system configuration of a strange
 star.
\par
 In order to minimize the drag energy, all the superconducting species have to
 share common normal cores. We call such a configuration a {\em vortex bundle}.
 The circulation of the $j$-th superconducting species around a circle of
 radius $r$ centered at the axis of rotation of the star in the co-rotating
 frame is given by
\begin{eqnarray}
 0 & = & \oint {\bf v}_j \cdot d{\bf l} \nonumber \\ & = & \oint
 \frac{\hbar}{m_j^*} \nabla \varphi_i d{\bf l} + \frac{q_j^*}{m_j^* c} \int
 {\bf B} \cdot d{\bf S} - \int \nabla\times \left( {\bf\Omega}\times {\bf r}
 \right) \cdot d{\bf S} \nonumber \\ & = & \frac{h N_j L}{m_j^*} + \frac{q_j^*
 \Phi_v L}{m_j^* c} + \frac{q_j^* \pi r^2 B_g}{m_j^* c} - 2\pi r^2 \Omega
 \mbox{~,}
\end{eqnarray}
 where $\Omega$ is the angular speed of the strange star seen by an external
 inertial observer, $N_j$ is the number of vortex quantum per bundle for
 species $j$, $L$ is the number of vortex bundles, $\Phi_v$ is the magnetic
 flux in the core of a vortex bundle, and $B_g$ is the global uniform (London)
 magnetic field in the star. Thus, the vortex bundle density ${\cal D}$ is
 given by
\begin{equation}
 {\cal D} \left( h N_j + \frac{q_j^* \Phi_v}{c} \right) = 2 m_j^* \Omega -
 \frac{q_j^* B_g}{c} \hspace{0.5in} \mbox{for all~} j. \label{E:Quant_1}
\end{equation}
\par\indent
 Using the same idea, we can show that the speed of superconducting species $j$
 at a small distance $r$ from the core of a vortex bundle as seen in the
 co-rotating frame is given by
\begin{equation}
 v_j (r) = \frac{1}{2\pi m_j^* r} \left[ h N_j + \frac{q_j^* \Phi_v}{c} \right]
 + \frac{q_j^* B_g r}{2m_j^* c} - \Omega r \hspace{0.5in} \mbox{for all~} j.
 \label{E:Quant_2}
\end{equation}
 Thus, the co-moving requirement of super-currents at all spatial points
 requires that (a) $q_j^* B_g / 2m_j^* c - \Omega$ is a constant for all $j$,
 which is possible only if the global uniform London field $B_g$ is zero; and
 (b) $h N_j / m_j^*  + q_j^* \Phi_v / m_j^* c$ is a constant for all $j$,
 implying that the integers $N_j$, and the magnetic flux $\Phi_v$ are chosen
 in such a way that
\begin{equation}
 \frac{h N_j}{m_j^*} + \frac{q_j^* \Phi_v}{m_j^* c} = K \label{E:Strong_Limit}
\end{equation}
 for some constant $K \neq 0$ for all $j$. $K$ can be interpreted as the
 circulation of a vortex bundle in this strongly interacting superconducting
 system.
\par
 The necessary and sufficient conditions for the existence of solution in
 Eq.~(\ref{E:Strong_Limit}) are proven in Appendix~\ref{S:App}. It turns out
 that the co-moving constraint (Eq.~(\ref{E:Strong_Limit})) is very stringent:
 In general, solution may not exist for a system involving more than two
 species of superconducting Cooper pairs. And in the case the solution of
 Eq.~(\ref{E:Strong_Limit}) does not exist, the only way out is that
 superconductivity in some species are destroyed. Luckily, in case of the
 strange star matter, there is only two species of Cooper pairs, namely, ud-du
 and ss, so that the solution in the strongly interacting limit exists. In this
 limit, we expect {\em all} the quark flavors to form quantized vortices when
 the star rotates. The normal vortex core of each superconducting species
 shares a common region of space, in the form of vortex bundles. Magnetic field
 may be present in the vortex bundle cores. Moreover, stellar rotation is made
 possible by the formation of vortex bundles rather than by a uniform London
 magnetic field.
\par
 In case of the strange star matter, the ud-du and ss Cooper pairs have charges
 $q_{ud}^* = e/3$ and $q_s^* = -2e/3$, respectively. Also, the effective mass
 of a strange quark at the strange star interior $m_s \sim 175$~MeV
 (\cite{Strange_Pulsar_EOS}). Therefore the effective mass of an ss Cooper
 pair, $m_s^*$, is approximately 350~MeV. The effective mass of a u or d quark,
 $m_{ud}$, is of order of 10~MeV, giving $m_{ud}^* \sim 20$~MeV. Since
 $(m_{ud}, q_{ud})$ and $(m_s, q_s)$ are linear independent, the solution of
 Eq.~(\ref{E:Strong_Limit}) is given by (see Appendix~\ref{S:App})
\begin{equation}
 \left( \begin{array}{c} K \\ \Phi_v \end{array} \right) = \frac{h}{[m_s^*
 q_{ud}^* - m_{ud}^* q_s^*]} \left( \begin{array}{c} q_{ud}^* N_s - q_s^*
 N_{ud} \\ c [ m_{ud}^* N_s - m_s^* N_{ud} ] \end{array} \right) \mbox{~,}
 \label{E:Solut} 
\end{equation}
 where $N_{ud}$ and $N_s$ are the number of quanta per vortex bundle for ud-du
 and ss Cooper pairs respectively, and $K \neq 0$. In the zero temperature
 limit, the system will choose the ground state configuration out of the above
 infinitely many solutions. Similar to Eq.~(\ref{E:F_1}), the Ginzburg-Landau
 free energy per unit volume in the strongly interacting limit is given by
\begin{equation}
 \frac{F}{V} \approx \left\{ \sum_j \left[ \frac{h^2 \rho_{sj} N_j^2}{4\pi
 {m_j^*}^2} \,\left[ \ln \left( \frac{R_{cj}}{\xi_j} \right) - \frac{3}{4}
 \right] + \frac{\pi \xi_j^2}{2} N_j(0) \Delta_j^2 \right] + \frac{\Phi_v^2}{8
 \pi^2 \lambda^2} \right\} \,| {\cal D} | \mbox{~,} \label{E:F_mi_alt}
\end{equation}
 where $\lambda$ is the penetration depth, and $\cal D$ is the number density
 of vortex bundles. From Eq.~(\ref{E:Quant_1}), ${\cal D} ( h N_j + q_j^*
 \Phi_v / c ) = 2m_j^* \Omega$ in the strongly interacting limit. Consequently,
\begin{eqnarray}
 \frac{F}{V} & \approx & \left\{ \sum_j \frac{1}{h N_j c + q_j^* \Phi_v} \left[
 \frac{h^2 \rho_{sj} N_j^2}{2\pi m_j^*} \,\left[ \ln \left(
 \frac{R_{cj}}{\xi_j} \right) - \frac{3}{4} \right] + \pi m_j^* \xi_j^2 N_j(0)
 \Delta_j^2 \right] + \right. \nonumber \\ \nonumber \\ & & \hspace{0.5in}
 \left. \frac{\Phi_v^2 m_1^*}{4\pi^2 \lambda^2 (h N_1 c + q_1^* \Phi_v)}
 \right\} \,c |\Omega | \label{E:F_mi}
\end{eqnarray}
 increases (approximately) linearly with the angular speed $|\Omega |$ of the
 star.
\par
 Now, we estimate the average free energy density for strange star matter. BCS
 theory tells us that $\Delta = 1.76 k T_c$ for s-wave paired superconductor.
 Putting $k T_c \sim 400$~keV for strange stellar matter (\cite{Bailin84}), we
 find $\Delta_{ud} \sim 700$~keV. The number densities for u and d quarks are
 given by (\cite{Alcock86})
\begin{equation}
 n_j = \frac{1}{\pi^2} \left( 1 - \frac{2\alpha_s}{\pi} \right) \mu_j^3
 \hspace{0.5in} \mbox{for~} j = \mbox{u,d,} \label{E:quark_num_den}
\end{equation}
 where $\mu_j$ is the chemical potential. Putting $\alpha_s = 0.5$ and $\mu_u
 \approx \mu_d \sim 400$~MeV (\cite{Strange_Pulsar_EOS}), $n_u \approx n_d \sim
 5.7 \times 10^{38}$~cm$^{-3}$. Since the electron number density is much less
 than that of the quarks (\cite{Bailin84}; \cite{Alcock86};
 \cite{Strange_Pulsar_EOS}), the charge neutrality condition reads
\begin{equation}
 \frac{2}{3} n_u - \frac{1}{3} n_d - \frac{1}{3} n_s = n_e \approx 0 \mbox{~.}
\end{equation}
 Therefore, we expect $n_s \approx n_d$.
\par
 For both u and d quarks, $E_F \approx \mu \sim 400$~MeV (\cite{Bailin84}), and
 hence from Eq.~(\ref{E:N0_def}, $N_{ud} (0) \sim 6.6\times
 10^{41}$~erg$^{-1}$\,cm$^{-3}$. Also, $m_u \approx m_d \sim 10$~MeV. The
 coherence length $\xi_j$ can be estimated in the framework of relativistic
 Ginzburg-Landau theory. Bailin and Love (1984) argue that
\begin{equation}
 \xi_j^3 \approx \frac{7\zeta (3)}{2 kT_{c_j} p_{F_j} \mu_j} \mbox{~,}
 \label{E:xi}
\end{equation}
 where $\zeta (3) \sim 1.20$, and $p_{F_j}$ is the Fermi momentum of species
 $j$ which is approximately equal to $E_{F_j} / c$ inside a strange star. Thus,
 $\xi_{ud} \sim 8.0$~fm. Since the quark superconductor is likely to be
 marginally type~I, $\lambda \approx \sqrt{2} \xi_{ud} \sim 11.3$~fm. Assuming
 $\ln ( R_{cj} / \xi_j )$ to be of order of 10, I numerically compute the
 average free energy density in Eq.~(\ref{E:F_mi}) for all possible values of
 $N_{ud}$ and $N_s$. I find that the ground state configuration for strange
 star matter is achieved when $N_s = 1$, $N_{ud} = 0$. Consequently, the values
 of $K$ and $\Phi_v$ in the ground state configuration are given by
\begin{mathletters}
\begin{equation}
 K = \frac{h}{m_s^* - 2m_{ud}^*} \sim 0.012\mbox{~cm}^2\,\mbox{s}^{-1}
 \label{E:SS_K}
\end{equation}
and
\begin{equation}
 \Phi_v = \frac{3hc m_{ud}^*}{e (m_s^* - 2m_{ud}^*)} \sim 8.0\times 10^{-8}
 \mbox{~G}\,\mbox{cm}^2 \label{E:SS_Phi}
\end{equation}
\end{mathletters}
\par\noindent
 respectively.
\par\indent
 In summary, the star behaves quite differently in the non-interacting and
 strongly interacting limits. In the non-interacting limit, a global magnetic
 field is set up, and vortices are formed in all but one superconducting
 species when the system rotates. In contrast, vortex bundles in the form of
 Abrikosov lattice is present in a rotating strongly interacting system.
 Moreover, the global uniform London magnetic field is no longer present.
 Regarding the coupling between the quarks as a tuning parameter, it is
 interesting to map out the phase diagram of this system. And this will be
 carried out in future works.
\section{Vortex Inter-pinning And Its Astrophysical Consequences \label{S:Ap}}
 The formation of vortices in a rotating superconducting strange star has at
 least three possible observational consequences if one assumes that the
 observed pulsars are in fact strange stars instead of neutron stars. First,
 the heat capacity of the star and the cooling processes are modified and its
 effect on the strange star cooling has been studied (\cite{SS_Cooling};
 \cite{D_Page}). In this section, we concentrate on the second effect, namely,
 the magnetic field decay due to inter-pinning of vortex bundles and the
 magnetic fluxoids. I shall discuss briefly possible magnetic field alignment
 due to inter-pinning in \S\ref{S:Forces} as well.
\par
 Nucleation energy is required in the creation of normal cores in both vortex
 bundles and magnetic fluxoids in a strange star. Therefore, it is
 energetically favorable for a vortex bundle to ``pin'' to a magnetic fluxoid
 so that a lesser volume of normal strange matter has to be nucleated. Thus,
 the dynamics of vortex bundles and magnetic flux tubes are coupled together.
 Similar to the case of superfluid neutron vortices in a rotating neutron star
 (\cite{VCM}), Eq.~(\ref{E:F_mi}) tells us that in the strongly interacting
 limit, the quark vortex bundle density is directly proportional to the angular
 speed $|\Omega |$ of the rotating strange star. As a strange star spins down,
 its vortex bundles in the star core have to move radially outward from the
 rotational axis (and eventually some of them will annihilate near the stellar
 surface). The pinning of quark vortex bundles and magnetic flux tubes implies
 that the rotational and magnetic evolution of a strange star are coupled.
\par
 Alternatively, the vortex bundle and the flux tubes can pin together by
 interaction of their core magnetic fields. When two magnetic carrying wires
 are placed together, an energy change of ${\bf B}_1 \cdot {\bf B}_2 V / 4\pi$
 is expected, where $V$ is their interacting volume, and ${\bf B}_j$ are their
 magnetic field strength. Depending on their relative field orientation, the
 two wires experience either mutual attraction or repulsion. In the former
 case, the wires will pin to each other; and in the latter case, the wires will
 avoid each other, and hence effectively ``pin'' to the inter-wire spaces. In
 fact, similar ideas have been applied to the proton fluxoids and magnetic
 vortices in neutron star cores; and their possible observational consequences
 have been studied (\cite{Sauls_Review}; \cite{Srinivasan}; \cite{Chau92};
 \cite{Ding93}).
\subsection{Estimation Of Pinning Energy Per Intersection \label{S:E_P}}
 Now we estimate the pinning energy per intersection of a vortex bundle with a
 flux tube by both pinning mechanisms. The pinning energy due to nucleation per
 intersection is given by
\begin{equation}
 \left( E_p \right)_{\rm nucl} \approx \frac{1}{2} \sum_j N_j(0) \Delta_j^2 V_j
 \mbox{~.} \label{E:Pin1}
\end{equation}
 The intersection volume for species $j$, $V_j$, is given by
\begin{equation}
 V_j \approx \alpha \pi {\cal N}_{\rm flux}^{1/2} \xi_j^3 \mbox{~,}
 \label{E:Volume}
\end{equation}
 where $\xi_j$ is the coherence length for species $j$, ${\cal N}_{\rm flux}$
 is the number of flux quantum in a flux tube, and $\alpha$ is a geometrical
 factor depending on the angle between the magnetic flux tube and the vortex,
 $\theta$, and their elastic moduli. For stiff magnetic flux tubes and vortex
 bundles, $\alpha \approx 2\,\mbox{cosec}\,|\theta |$.
\par
 We estimate the nucleation pinning energy as follows: for an s-wave
 superconductor, BCS theory tells us that $\Delta = 1.76 k T_c$; and for a
 p-wave superconductor, $\Delta = 2.4 k T_c$ (\cite{Baym75}). Putting $k T_c
 \sim 400$~keV for strange stellar matter (\cite{Bailin84}), we obtain
 $\Delta_{ud} \sim 700$~keV and $\Delta_s \sim 960$~keV. For ud-du quark Cooper
 pairs, $E_F \approx \mu \sim 400$~MeV (\cite{Bailin84}). And beta equilibrium
 requires $E_{F_s} \approx \mu_s = \mu_d \approx 400$~MeV. The coherence
 lengths are given by Eq.~(\ref{E:xi}), giving us $\xi_{ud} \approx \xi_s \sim
 8$~fm. So from Eqs.~(\ref{E:N0_def}),~(\ref{E:Pin1}) and~(\ref{E:Volume}),
 we obtain
\begin{equation}
 \left( E_p \right)_{\rm nucl} \sim 2.4 {\cal N}_{\rm flux}^{1/2}\,
 \mbox{cosec}\,|\theta | \mbox{~MeV.} \label{E:Pin2}
\end{equation}
\par\indent
 Similarly, the magnetic pinning energy is given by (\cite{Chau92};
 \cite{Ding93})
\begin{equation}
 \left( E_p \right)_{\rm mag} \approx \frac{\left| {\bf B}_{\rm vortex} \cdot
 {\bf B}_{\rm fluxoid} \right| }{4\pi} \,V \approx \frac{B_c \Phi_v \lambda
 {\cal N}_{\rm flux}^{1/2}}{2\pi} | \cot \theta | \mbox{~,} \label{E:Pin_Mag}
\end{equation}
 where $B_c$ is the critical magnetic field. Since the flux quantum for a
 strange star flux tube equals $hc / 2 (2e/3) = 3hc / 4e$, and $\lambda \sim
 11.3$~fm, we obtain $B_c \sim 7.7\times 10^{16}$~G. By putting the value of
 $\Phi_v$ obtained in Eq.~(\ref{E:SS_Phi}), we find 
\begin{equation}
 \left( E_p \right)_{\rm mag} \sim 690 {\cal N}_{\rm flux}^{1/2} | \cot \theta
 | \mbox{~MeV.} \label{E:Pin_Mag_2}
\end{equation}
 So, unless the magnetic flux tubes and the vortices are perpendicular to each
 other (this happens when the rotational and magnetic axes of the star are
 orthogonal to each other), the magnetic pinning energy is approximately two
 order of magnitude larger than the nucleation pinning energy. Therefore, we
 shall use the magnetic pinning energy in our subsequent calculations.
\subsection{Forces Acting On Flux Tubes And Vortices \label{S:Forces}}
 Following Ding et al. (1993), the major forces acting on flux tubes and
 vortices of a spinning down strange star are summarized below:
\newcounter{n_force}
\begin{list} {\noindent(\alph{n_force})}{\usecounter{n_force}
 \settowidth{\labelwidth}{\rm (a)} \settowidth{\labelsep}{\rm ~\,}
 \setlength{\leftmargin}{\labelwidth} \addtolength{\leftmargin}{\labelsep}
 \setlength{\rightmargin}{0in}}
\item {\em Inter-pinning force:} Since it is energetically favorable for the
 magnetic flux tubes and the vortex bundles to pin with each other, a force of
 order of $E_p / \xi$ is experienced by both parties when one tries to pull
 them apart. (Similar assertion is true when the magnetic pinning force is
 repulsive.) This inter-pinning force encourages the magnetic flux tubes and
 vortex bundles to move together.
\item {\em Thermal activation:} Since $E_p$ is of order of 100~MeV and the
 interior temperature of the star is $\lesssim 10$~keV, random thermal noise
 can be an important source of ``force''. In particular, thermal activation can
 depin a vortex with a flux line.
\item {\em Magnus force:} The magnetic flux tubes couple to the thin crust of
 the strange star via electromagnetic interaction between its strong core
 magnetic field and the background electron plasma. Therefore, the flux tubes
 co-rotate, and hence spin down, with the crust of the star. The vortex bundle
 core, due to inter-pinning, may spin down with the crust as well.
 Eq.~(\ref{E:Quant_1}) tells us that the only way to slow down the angular
 speed of a rotating superconducting quark fluid is by reducing its vortex
 bundle density. This can in turn be achieved only by moving the vortex bundles
 radially outward from the axis of rotation of the star. So as the star
 slows down, a steady angular velocity difference between the superconducting
 quark fluid and the crust is developed. Consequently, a hydrodynamic force,
 called the Magnus force, acting on the vortex bundles is developed. Magnus
 force tries to push the vortex bundle radially outward from the rotational
 axis of the star, thereby reducing the angular velocity difference between the
 superconducting quark fluid and the crust. The combined effect of the Magnus
 force and the inter-pinning force may cause a spin down induced magnetic field
 expulsion in a strange star. The Magnus force also defines a preferred
 direction for the thermal activation (i.e., thermally assisted creeping).
 \par
 For a inter-winded network of pinned magnetic flux tubes and quark vortex
 bundles, the average force per unit length acting on a magnetic flux tube due
 to the Magnus force acting on the vortex bundles is given by (compare with
 \cite{Ding93})
 \begin{equation}
  {\bf f}_{f,\rm Mag} \approx \frac{L_v}{L_f} \,\rho K r \omega \hat{\bf e}_r
  \mbox{~,} \label{E:Magnus}
 \end{equation}
 where $r$ is the distance from the rotational axis, $\omega$ is the angular
 velocity difference between the superconductor ($\Omega_s$) and the stellar
 crust ($\Omega_c$), $K$ is the circulation given by Eq.~(\ref{E:SS_K}), $\rho$
 is the total matter density, $L_f$ and $L_v$ are the total number of flux
 tubes and vortex bundles respectively in the star. In fact, $L_f$ and $L_v$
 can be determined by considering the total magnetic flux and the circulation
 of superconducting current of the star. They are given by
 \begin{equation}
  L_f = \frac{4\pi e R_{\rm star}^2 | B_{\rm star} |}{3hc {\cal N}_{\rm flux}}
  \mbox{~,} \label{E:L_f_def}
 \end{equation}
 and
 \begin{equation}
  L_v = \frac{2\pi R_{\rm star}^2 |\Omega |}{K} \equiv \frac{4\pi
  R_{\rm star}^2 (m_s - 2m_{ud})}{h} \,|\Omega | \mbox{~,} \label{E:L_v_def}
 \end{equation}
 where $R_{\rm star}$ and $B_{\rm star}$ are the radius and total magnetic
 field strength of the star respectively.
 \par
 The angular velocity difference, $\omega$, is likely to remain at its steady
 state value, which can be deduced from the vortex creep theory (\cite{VCM84};
 \cite{VCM}) and is given by
 \begin{eqnarray}
  \omega_\infty & = & \pm \frac{kT}{\rho K b_p r \lambda} \sinh^{-1} \left[
  \frac{r}{4\Omega v_0} |\dot\Omega | \exp \left( \frac{E_p}{kT} \right)
  \right] \nonumber \\ & \approx & \pm \frac{E_p}{\lambda\rho K b_p r}
  \nonumber \\ & \approx & \pm \frac{B_c \Phi_v |\cos\theta |}{\pi\rho K r}
  \,\left( \frac{e | B_{\rm star} |}{3hc} \right)^{1/2} = \pm \omega_{cr}
  \mbox{~,}
 \end{eqnarray}
 where we have assumed that $E_p \gg kT$. Here $v_0 \sim 10^{13}$~cm\,s$^{-1}$
 is the microscopic creeping speed, $b_p \approx ( 1 / |\sin \theta | ) \times
 ( \pi R_{\rm star}^2 / L_f )^{1/2}$ is the mean distance between successive
 pinning sites along a vortex bundle, and $\omega_{cr}$ is the critical angular
 velocity lag above which it is no longer energetically favorable for the
 vortices and fluxoids to pin together. The plus (or minus) sign is taken if
 the outward moving speed of vortex bundles is greater (or less) than that of
 magnetic flux tubes. Therefore, the maximum possible average force per unit
 length vortex bundles can exert on a flux tube equals $\rho K R_{\rm star}
 \omega_{cr} L_v / L_f$.
 \par
 Finally, if the spin down of the star is mainly due to its dipole radiation
 lost, then the outward moving speed of the vortex bundles is given by
 (\cite{Ding93})
 \begin{equation}
  v_v (t) \approx \frac{B_{\rm star}^2 R_{\rm star}^7 \Omega^2 \sin^2\theta}{3I
  c^3} \mbox{~,} \label{E:v_v}
 \end{equation}
 where $I$ is the moment of inertia of the star.
\item {\em Buoyancy force:} The presence of magnetic stress in the core of a
 flux line decreases the internal density of quark matter (\cite{Muslimov85};
 \cite{Harvey}; \cite{Jones87}). Thus, the buoyancy force per unit length
 experienced by a flux line is
 \begin{equation}
  {\bf f}_{\rm buoy} = \frac{9 h^2 c^2 {\cal N}_{\rm flux} g}{128 e^2 \pi^2
  \lambda^2 c_s^2} \,\hat{\bf e}_r \approx \frac{9 h^2 c^2
  {\cal N}_{\rm flux}}{128 e^2 \pi^2 \lambda^2 R_{\rm star}} \,\hat{\bf e}_r
  \mbox{~,} \label{E:f_buoy}
 \end{equation}
 where $g$ is the local acceleration due to gravity, $c_s^2 \equiv dP / d\rho
 \approx g R_{\rm star}$ is the squared sound speed.
\item {\em Tension:} The combined pushing of all the vortex bundles pinned onto
 a given magnetic flux line may globally bend the flux line. In effect, a
 tension force is developed which tries to resist further deformation. The
 average tension per unit length is given by (\cite{Harvey})
 \begin{equation}
  {\bf f}_{\rm tens} = - \frac{9 h^2 c^2 {\cal N}_{\rm flux}}{256 e^2 \pi^2
  \lambda^2} \ln \left( \frac{\lambda}{\xi} \right) \frac{1}{s_c}
  \,\hat{\bf e}_r \approx - \frac{R_{\rm star} f_{\rm buoy}}{2 s_c} \ln \left(
  \frac{\lambda}{\xi} \right) \,\hat{\bf e}_r \mbox{~,} \label{E:f_tens}
 \end{equation}
 where $s_c$ is the radius of curvature of the flux tube.
\item {\em Electron drag force:} As a flux tube drifts out of the core, it will
 experience a drag force arising from the scattering of the degenerate
 relativistic electron. This drag force limits the speed of the flux tubes and
 hence the rate of magnetic field decay. In the absence of clumping of flux
 tubes, the drag force per unit length is given by\footnote{Note that there is
 an extra factor of 9/4 because the flux quantum for proton superconductor and
 strange matter superconductor are different. In addition, the case of flux
 clumping will be addressed in \S\ref{S:Decay_Num}.} (\cite{Muslimov85};
 \cite{Harvey}; \cite{Jones87})
 \begin{equation}
  {\bf f}_{\rm drag} = - \frac{27\pi}{1024} \,\frac{n_e h^2 c
  {\cal N}_{\rm flux}^{3/2} v_f}{E_f(e) \lambda} \,\hat{\bf e}_r \mbox{~,}
  \label{E:f_drag_force}
 \end{equation}
 where $E_f(e)$ is the electron Fermi energy, $v_f$ is the velocity of the flux
 tube, and $n_e$ is the electron number density and is given by
 (\cite{Alcock86})
 \begin{equation}
  n_e = \frac{\mu_e^3}{3\pi^2} \mbox{~.} \label{E:num_elec}
 \end{equation}
 The electron chemical potential in a strange star is about 20~MeV
 (\cite{Strange_Pulsar_EOS}), giving us $n_e \sim 3.5\times 10^{34}$~cm$^{-3}$.
 Thus, the electron number density is some two orders of magnitude smaller than
 that of a neutron star. Consequently, the electron drag force in a strange
 star is much weaker than that in a neutron star.
\end{list}
\par\indent
 The average radial velocity of a flux tube is approximately equal to its
 steady state radial velocity $v_f$, which can be calculated from the force
 balance equation
\begin{equation}
 {\bf f}_{\rm f,Mag} + {\bf f}_{\rm buoy} + {\bf f}_{\rm tens} +
 {\bf f}_{\rm drag} = {\bf 0} \mbox{~.} \label{E:F_Balance}
\end{equation}
 After finding $v_f$, the magnetic field $B_{\rm star}$ of the star can be
 computed by solving the equation (compare with the $\dot{\Omega}$ equation in
 \cite{VCM84})
\begin{equation}
 \dot{B}_{\rm star} = - \frac{2B_{\rm star} v_f}{R_{\rm star}} \mbox{~.}
\end{equation}
\subsection{The Alignment Torque \label{S:Alignment_Torque}}
 In general, magnetic flux tubes incline at a non-zero angle $\theta$ to the
 rotational axis of the star. Thus, the magnitude (and sometimes even the
 direction) of force acting on a flux tube by the vortex bundles changes as
 we go along the flux tube itself. The combined forces acting on a magnetic
 flux tube due to vortex bundles, therefore, produce a total force together
 with a torque (Ruderman~1991abc). While the total force may lead to a magnetic
 field decay in a spinning down strange star, the torque tends to push the
 field lines towards the equator as the star spins down. Similarly, it tends to
 push the field lines towards the poles as the star spins up (\cite{Align2}).
\par
 The effect of this torque on the time evolution of magnetic field is not
 completely clear. If the crustal material of a strange star, which makes up of
 nuclear instead of strange matter, is strong enough to support a shear, then
 the direction of magnetic field will not change with time. On the other hand,
 if the crustal nuclear material is weak and brittle, direction of magnetic
 field may change via a series of ``crust cracking'' (Ruderman~1991abc). A more
 complete discussion on the effect of this alignment torque will be reported in
 future works.
\subsection{The Crustal Magnetic Field \label{S:Stru}}
 So far, I am concentrating on the magnetic field in the core of the star. In
 this subsection, I show that the presence of crustal magnetic field does not
 seriously affect the magnetic field evolution of a strange star. The density
 of the strange stellar crust must be below the neutron drip density ($\sim 4.3
 \times 10^{11}$~gm\,cm$^{-3}$), otherwise the dripped neutrons can convert
 into strange matter by making contact with them. Since the most favorable
 nucleus just below neutron drip is $^{118}$Kr (\cite{Baym75}), spacing between
 the nuclear lattice just below the neutron drip $b$ is about 70~fm. Hence, the
 Young's modulus of this lattice is about $(Ze)^2 / b^4 \sim
 10^{29}$~dyn\,cm$^{-2}$. (The Young's modulus of the nuclear lattice is much
 smaller if it is not prefect or if the density of the crust is lower than for
 the neutron drip.) If the magnetic flux is completely expelled from the core,
 the inner layer of the normal matter crust has to sustain a stress of $\approx
 B^2 \Delta R / 4\pi$ per unit length, where $\Delta R \sim 100$~m is the
 thickness of the crust. The magnetic stress lengthens the nuclear lattice;
 however, such a lengthening must not be greater than the $\lesssim 10$~fm 
 electro-static gap between strange and nuclear matter in the star. Otherwise,
 strange matter conversion will take place and the crust will be destroyed.
 After some computation, I find that the maximum field and flux the nuclear
 crust can sustain are $\sim 10^7$~G and $\sim 7\times 10^{17}$~G\,cm$^2$
 respectively. This is much smaller than the initial magnetic flux of a
 strange pulsar. So we have two possibilities: (a) there is an efficient field
 decay mechanism operating in the thin nuclear matter crust; or (2) the crust
 breaks at some point when majority of the field in the core is expelled,
 eventually making the star a bare strange matter object. In either case, the
 contribution of the magnetic field in the crust does not play an important
 role in the magnetic evolution of the entire star. Thus, we shall neglect the
 presence of crustal field in our subsequent analysis.
\subsection{Magnetic Field Decay --- Numerical Results \label{S:Decay_Num}}
 I numerically compute the magnetic field decay due to inter-pinning of
 magnetic flux tubes and vortex bundles. I assume that the initial magnetic
 field $B_{\rm star} (0)$ is uniformly distributed in the stellar core, and we
 also neglect the existence of a thin nuclear matter crust. The initial angular
 speed of the star is set to be $\Omega (0)$. The effect of the alignment
 torque is also ignored. To simplify calculations, we follow Ding et al. (1993)
 to combine the buoyancy force (Eq.~(\ref{E:f_buoy})) and the tension force
 terms (Eq.~(\ref{E:f_tens})) because of their similar dependence on
 parameters. We write
\begin{equation}
 f_{\rm buoy} + f_{\rm tens} = \gamma f_{\rm buoy} \mbox{~,} \label{E:Def_Gam}
\end{equation}
 where $\gamma = 1 - ( R_{\rm star} / 2 s_c ) \ln (\lambda / \xi )$ is a time
 dependent quantity of order of unity in most parts of the star
 (\cite{Harvey}).
\par
 What happens if the magnetic flux lines form a clump? Provided that the size
 of such a clump is smaller than the electron mean free path, electrons may
 collectively scatter with a number of flux lines within a clump leading to a
 dramatic change in the drag force (\cite{Mal}). In addition, clumping leads to
 an increase in the local density of pinning centers. Moreover, the force
 between flux tubes within a clump may be important. Thus, our mean field
 estimate of the force acting on the flux tubes by the vortex bundles will
 change as well.
\par 
 We model the clumping effect by a renormalization scheme. By coarse graining
 to the level of a clump, the behavior of the flux tubes inside a clump is
 similar to that of a single magnetic flux tube carrying the same amount of
 flux as the clump provided that the size of the clump is much smaller than the
 electron mean free path. In addition, the magnetic field strength in this
 single magnetic flux tube equals to that in each of the individual flux tubes
 in the clump. That is, we may replace ${\cal N}_{\rm flux}$ by
 ${\cal N}_{\rm flux} {\cal N}_{\rm clump}$. As a result, the electron drag
 force, vortex acting force, and the buoyancy force on a clump are given by
\begin{mathletters}
\begin{equation}
 f_{\rm drag,coll} \approx {\cal N}_{\rm clump}^{3/2} f_{\rm drag} \mbox{~,}
 \label{E:f_drag_force_coll}
\end{equation}
\begin{equation}
 f_{\rm Mag,coll} \approx {\cal N}_{\rm clump} f_{f,\rm Mag} \mbox{~,}
\end{equation}
 and
\begin{equation}
 f_{\rm buoy,coll} \approx {\cal N}_{\rm clump} \gamma f_{\rm buoy}
\end{equation}
\end{mathletters}
 respectively. So the electron drag force becomes dominant when clumping of
 flux lines is serious or when the quark superconductivity is extremely type~I.
 And from now on, the symbol ${\cal N}_{\rm flux}$ should be interpreted as the
 product of the number of flux quantum in each flux tube and the number of flux
 tubes in a clump.
\par
 Let us consider a 1.4M$_\odot$ strange star with (core) radius 11~km whose
 density in the outer region of the strange core is around $5\times
 10^{14}$~gm\,cm$^{-3}$. The moment of inertia of the star is about $1.6\times
 10^{45}$~gm\,cm$^2$ (\cite{Strange_Pulsar_EOS}). We also fix $\theta =
 45^\circ$. We take a typical star with initial magnetic field $B_{\rm star}
 (0) = 10^{12}$~G, initial angular speed $\Omega (0) = 2000$~rad\,s$^{-1}$ (and
 hence an initial period of about 3~ms), $\gamma = 0.5$, and
 ${\cal N}_{\rm flux} = 1$ as our ``reference'' star. Then the effects of the
 initial magnetic field, initial angular speed, the value of $\gamma$ in the
 effective buoyancy force, and the number of flux quanta in a flux clump
 ${\cal N}_{\rm flux}$ on the stellar magnetic evolution can be studied by
 varying these parameters one at a time.
\par
\begin{figure}
 \plotone{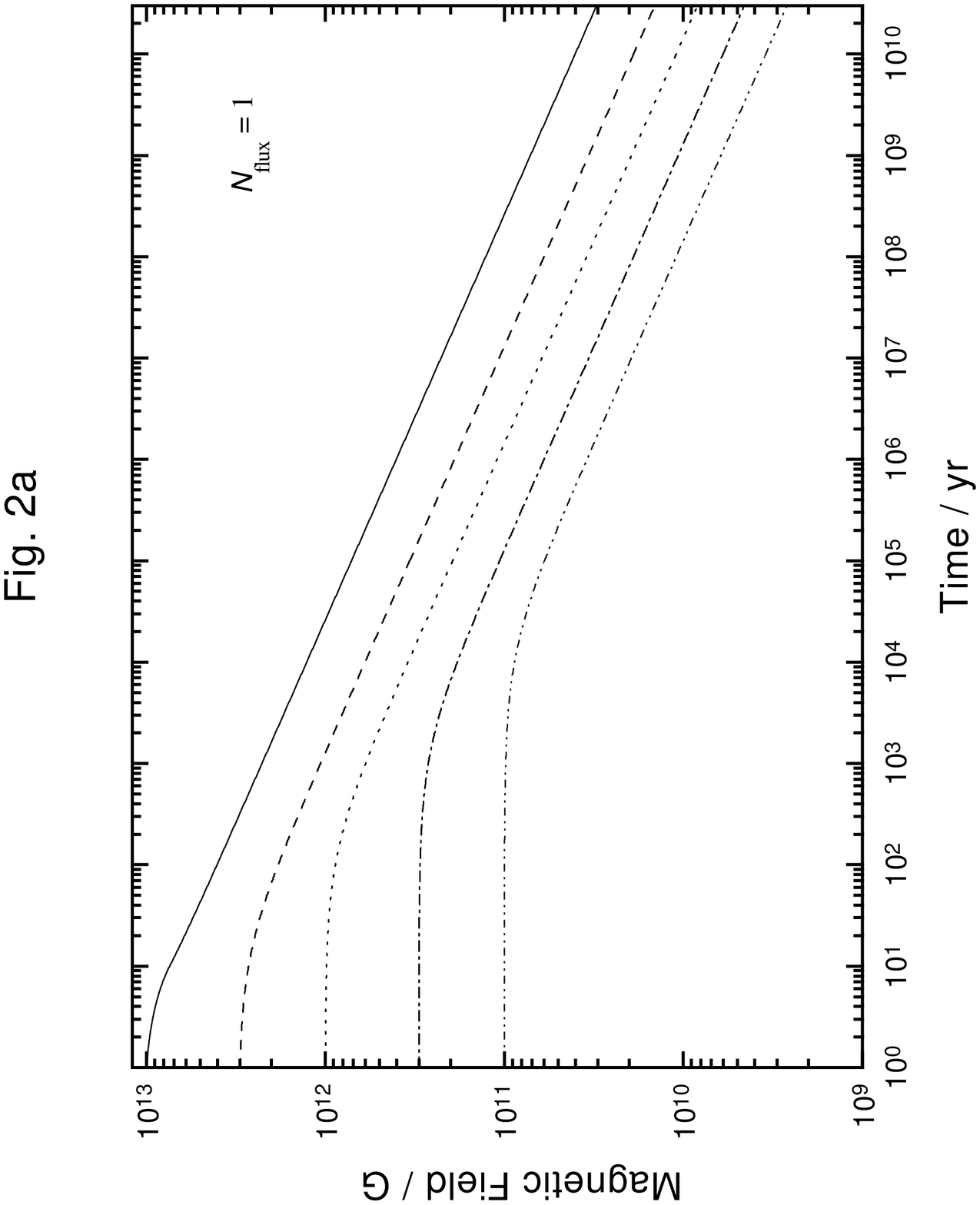}
 \caption{The magnetic field decay of a strange star as a function of its
  initial $B$ field. The initial field measured in G is varied from $10^{13}$
  (solid line), $3\times 10^{12}$ (dash line), $10^{12}$ (dotted line), $3
  \times 10^{11}$ (dash dotted line), to $10^{11}$ (short dash dotted line).
  (a) shows the field decay when ${\cal N}_{\rm flux} = 1$; while (b) shows the
  field decay when ${\cal N}_{\rm flux} = 10^6$. \label{F:Init_Field}}
\end{figure}
\begin{figure}
 \plotone{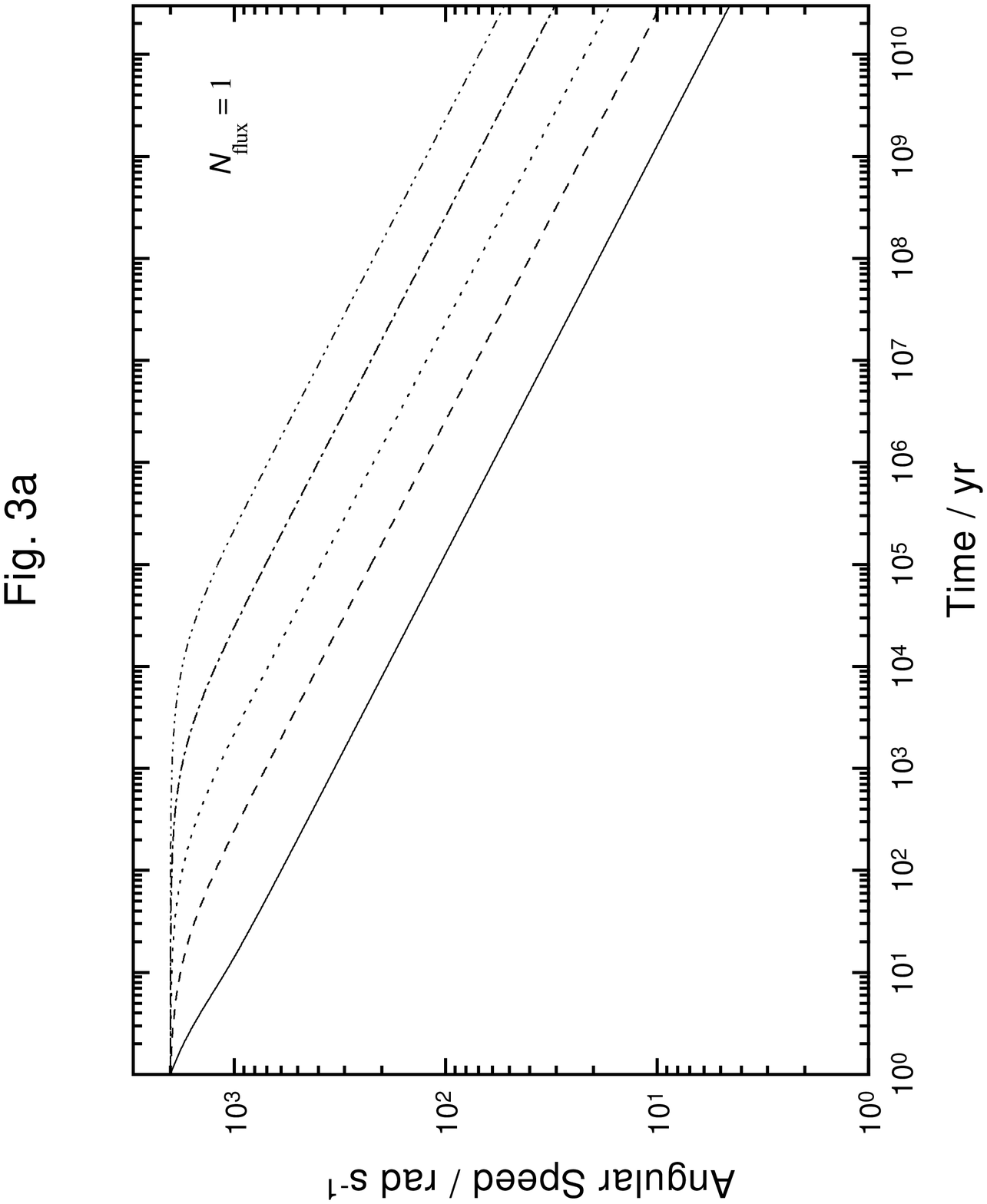}
 \caption{Spin down of a strange pulsar as a function of its initial magnetic
  field. The solid, dotted, dash, dash dotted, and short dash dotted lines
  correspond to initial magnetic fields of $10^{11}$, $3\times 10^{11}$,
  $10^{12}$, $3\times 10^{12}$ and $10^{13}$~G respectively. (a) and (b) are
  for ${\cal N}_{\rm flux} = 1$ and $10^6$ respectively. \label{F:Plot_Omega}}
\end{figure}
\begin{figure}
 \plotone{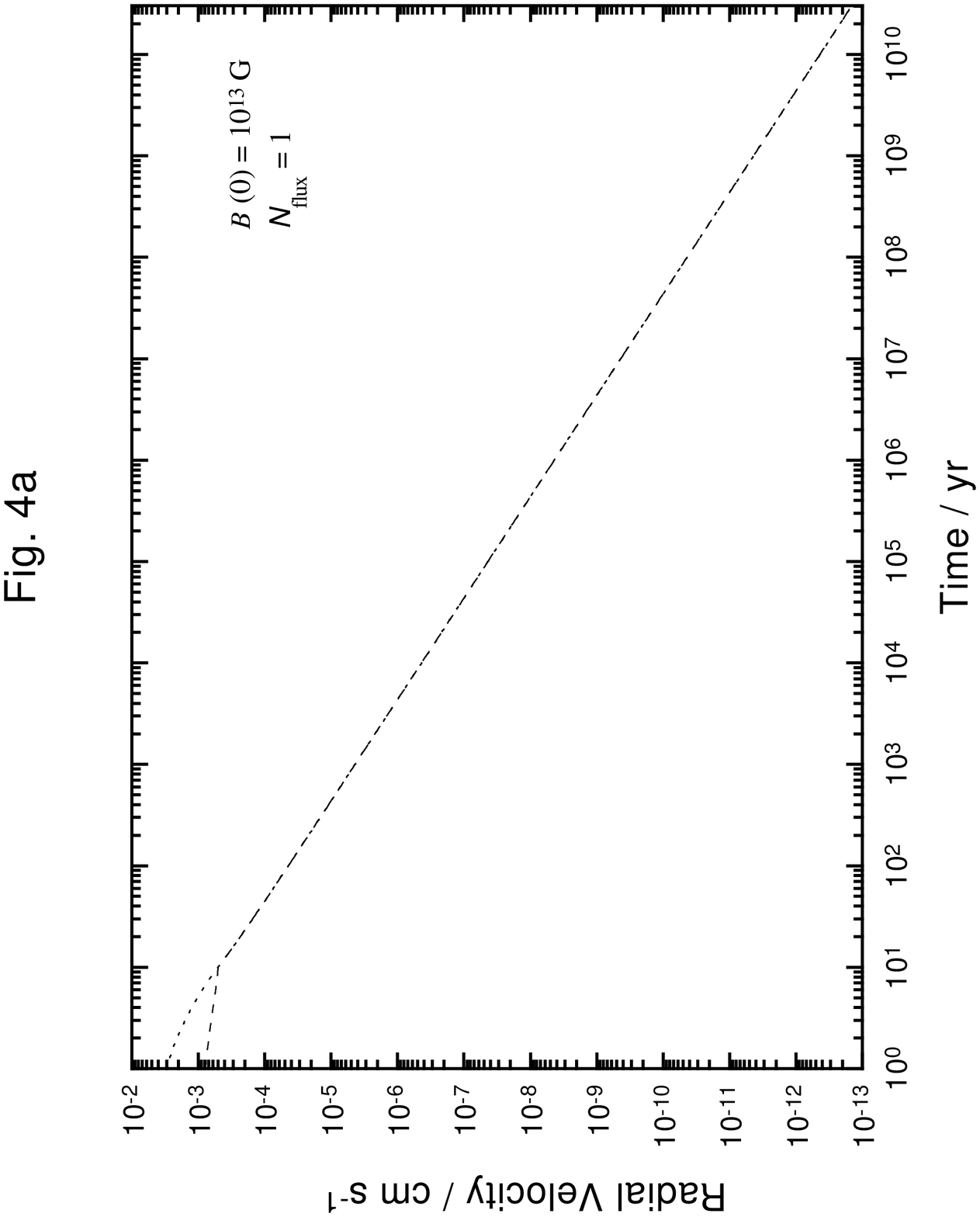}
 \caption{Radial velocities of vortex bundles (dotted line) and magnetic flux
  tubes (dash line) when (a) $B(0) = 10^{13}$~G, ${\cal N}_{\rm flux} = 1$; (b)
  $B(0) = 10^{12}$~G, ${\cal N}_{\rm flux} = 10^6$; (c) $B(0) = 3\times
  10^{12}$~G, ${\cal N}_{\rm flux} = 10^6$; and (d) $B(0) = 10^{13}$~G,
  ${\cal N}_{\rm flux} = 10^6$. The initial angular velocities in all four
  cases equal 2000~rad\,s$^{-1}$. \label{F:Velocities}}
\end{figure}

 We first investigate the case when there is no clumping, and the system is
 almost type~II. So we set ${\cal N}_{\rm flux} = 1$. As shown in
 Figs.~\ref{F:Init_Field}a and~\ref{F:Plot_Omega}a, both the magnetic field and
 angular speed remain approximately constant for a while which are followed by
 power law decays with exponents equal $-1/4$. Their transition times decrease
 with increasing initial field. In addition, the radial velocities of both the
 vortices and the fluxoids moves as a whole most of the time (see
 Fig.~\ref{F:Velocities}a).
\par
 We can explain both the field and angular speed evolution in a simple way.
 Since the initial magnetic field is not too high, the star only experiences a
 modest spin down. Thus, the vortex bundles are effectively pinned to the flux
 tubes (instead of thermally creep through them) leading to what Ding et al.
 (1993) called a ``co-moving phase''. In this phase, $B_{\rm star} (t) \sim
 \Omega (t)$ and hence the dipole spin down of the star is given by
\begin{equation}
 \dot\Omega = - \frac{\Omega^3 R_{\rm star}^6 B_{\rm star}^2 \sin^2\theta}{3I
 c^3} \approx - \frac{B_{\rm star}^2 (0) R_{\rm star}^6 \Omega^5 \sin^2
 \theta}{2I c^3 \Omega^2 (0)} \mbox{~.}
\end{equation}
 Upon integration, it is easy to show that
\begin{equation}
 \Omega (t) \approx \left[ \frac{4B_{\rm star}^2 (0) R_{\rm star}^6 t \sin^2
 \theta}{3Ic^3 \Omega^2 (0)} + \frac{1}{\Omega^4 (0)} \right]^{-1/4} \equiv
 \left[ \frac{t}{t_0} + \frac{1}{\Omega^4 (0)} \right]^{-1/4} \mbox{~.}
 \label{E:Weak_Field_Om}
\end{equation}
 Thus, $\Omega$ (and hence $B_{\rm star}$) is almost a constant when $t \ll
 t_0$ and $\Omega$ decays as a power law with an exponent $-1/4$ when $t \gg
 t_0$. This is consistent with the field and angular speed evolution we have
 plotted in Figs.~\ref{F:Init_Field}a and~\ref{F:Plot_Omega}a.
\par
 Now, we go on to consider the case when clumping is important, and when the
 quark superconductor is almost type~I. We illustrate the situation by putting
 ${\cal N}_{\rm flux} = 10^6$. As shown in Figs.~\ref{F:Init_Field}b
 and~\ref{F:Plot_Omega}b, when the initial field is low ($\lesssim 3\times
 10^{12}$~G), the magnetic and spin evolution behave almost in same way as in
 the case when ${\cal N}_{\rm flux} = 1$. The huge value of
 ${\cal N}_{\rm flux}$ implies that the electron drag force may be dominant. In
 order to push the flux tubes, the vortex bundles need to acquire a strong
 Magnus force by increasing the value of the angular velocity lag $\omega$.
 Nevertheless, $|\omega |$ cannot exceed its critical value $\omega_{cr}$. So,
 when the star spins down too quickly, the vortex bundles in the core have no
 choice but to thermally creep through the flux tubes leading to what Ding et
 al. (1993) called the ``forward creeping phase''. This picture is consistent
 with the numerical finding that forward creeping phase lengthens with
 increasing initial magnetic field (see Figs.~\ref{F:Velocities}b, c and~d).
 The small radial velocity of the flux tubes at this phase implies that
 $B_{\rm star}$ remains almost constant. It is until the onset of the co-moving
 phase due to a much slower spin down rate at later times that $B_{\rm star}$
 and $\Omega$ decay like $t^{-1/4}$ (see Figs.~\ref{F:Init_Field}b
 and~\ref{F:Plot_Omega}b).
\par
 Now we discuss a more interesting case when $B_{\rm star} (0) = 10^{13}$~G
 and ${\cal N}_{\rm flux} = 10^6$. Fig.~\ref{F:Init_Field}b shows that
 $B_{\rm star}$ decays exponentially with a characteristic time of about 5~Myr.
 However, at about 100~Myr, the decay stays almost constant for a while, and
 then it decays further like $t^{-1/3}$. Besides, Fig.~\ref{F:Plot_Omega}b
 shows that $\Omega$ decreases like a power law with an exponent $-1/2$ for
 almost 10~Myr. After that, $\Omega$ stays almost constant for a while which is
 then followed by a power law decay with an exponent about $-1/6$.
\par
 We can explain the behavior as follows: The rapid spin down due to high value
 of $B_{\rm star}$ leads to the forward creeping phase (see
 Fig.~\ref{F:Velocities}d). So,
\begin{equation}
 \dot\Omega \approx - \frac{\Omega^3 R_{\rm star}^6 B_{\rm star}^2 (0) \sin^2
 \theta}{3I c^3}
\end{equation}
 and hence
\begin{equation}
 \Omega (t) \approx \left[ \frac{2B_{\rm star}^2(0) R_{\rm star}^6 t \sin^2
 \theta}{3Ic^3} + \frac{1}{\Omega^2 (0)} \right]^{-1/2} \mbox{~.}
\end{equation}
 This accounts for the exponential decay in $\Omega$ starting from $3Ic^3 / 2
 B_{\rm star}^2 (0) R_{\rm star}^6 \Omega^2 (0) \sin^2\theta \approx 5$~yr.
 About $10^3$~yr or so, the star rotates so slowly that vortex bundle density
 becomes very low. At this moment, buoyancy becomes the dominant driving force
 for field decay and hence
\begin{equation}
 \dot{B}_{\rm star} = - \frac{2B_{\rm star} v_f}{R_{\rm star}} \approx -
 \frac{16\gamma E_f (e) c B_{\rm star}}{3\pi^3 e^2 n_e \lambda R_{\rm star}^2
 {\cal N}_{\rm flux}^{1/2}} \equiv - \frac{B_{\rm star}}{\tau_{\rm buoy}}
 \mbox{~.} \label{E:tau_buoy}
\end{equation}
 That is, $B_{\rm star}$ decays exponentially with a characteristic time
 $\tau_{\rm buoy}$ of about 4~Myr. Finally at time $\gtrsim 10$~Myr, dipole
 spin down of the star becomes very ineffective due to the small values of
 $\Omega$ and $B_{\rm star}$. At this time the radial velocity of flux tubes is
 faster than that of the vortex bundles and the star enters the ``reverse
 creeping'' phase (see Fig.~\ref{F:Velocities}d and \cite{Ding93}). The pinning
 force prevents the flux tubes from moving too fast and hence the rate of field
 decay is decreased. The force balance equation for the flux tubes reads
 $f_{\rm buoy,coll} - f_{\rm Mag,coll} = f_{\rm drag,coll} \approx 0$ (compare
 with Eq.~(\ref{E:F_Balance})). Since $f_{\rm buoy,coll}$ is time independent,
 $f_{\rm Mag,coll} \propto |\Omega \omega / B_{\rm star}|$, and $\omega \approx
 -\omega_{cr} \propto |B_{\rm star} |^{1/2}$, the force balance equation
 implies that $B_{\rm star} \sim \Omega^2$. As a result, $\Omega (t) \sim
 t^{-1/6}$ and $B_{\rm star} (t) \sim t^{-1/3}$ at the very late stage of the
 evolution of the star (see Figs.~\ref{F:Init_Field}b and~\ref{F:Plot_Omega}b).
\par
\begin{figure}
 \plotone{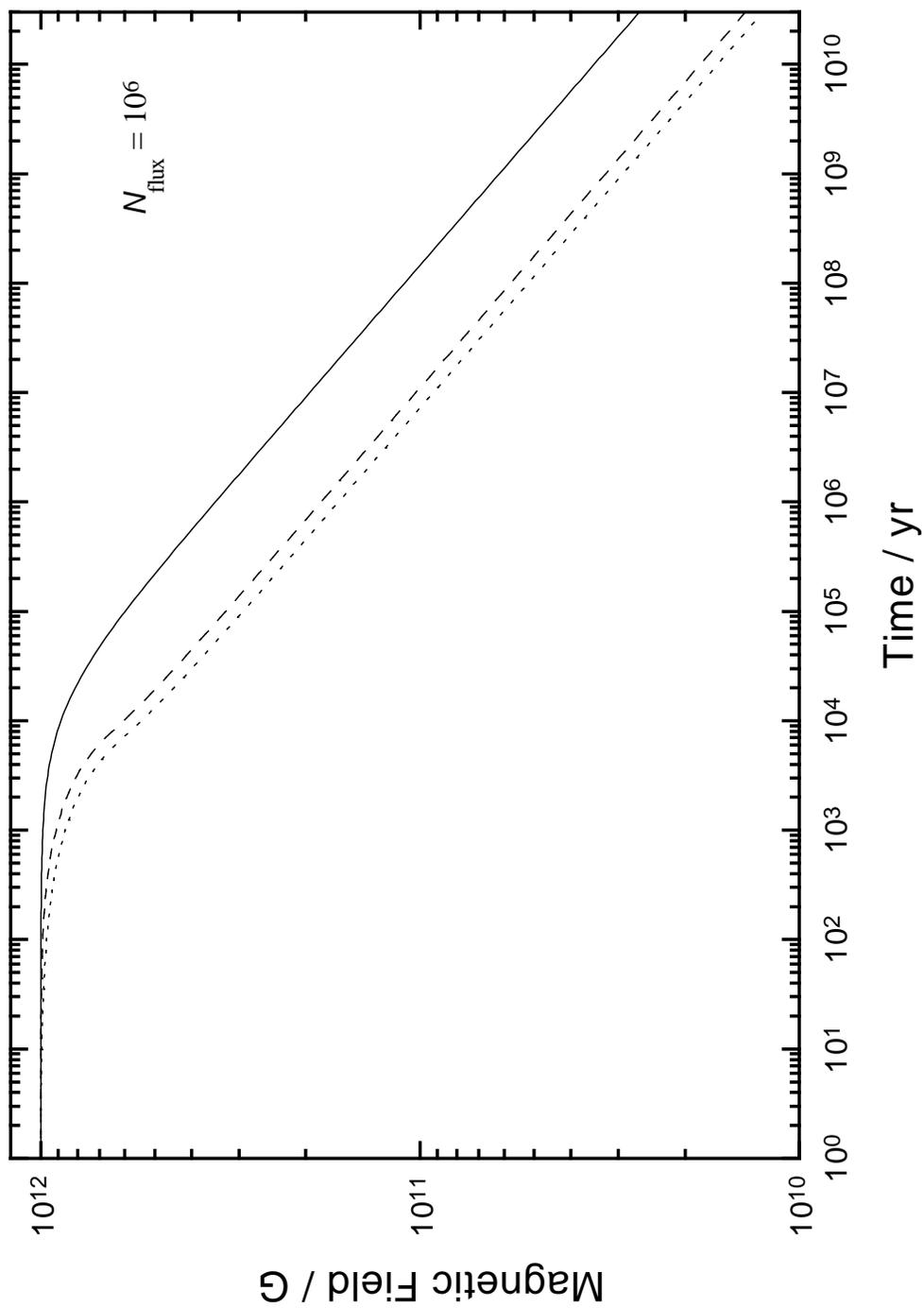}
 \caption{Magnetic field evolution as a function of initial angular speed. The
  solid, dash and dotted lines correspond to initial angular speed of 200,
  2000, and 12000~rad\,s$^{-1}$ respectively. ${\cal N}_{\rm flux}$ is set to
  $10^6$. \label{F:Omega}}
\end{figure}

 We now turn to the study of the effect of $\Omega (0)$ to the magnetic
 evolution of the star. Fig.~\ref{F:Omega} shows that the field decays more
 rapidly (due to the faster onset of the power law decay) as $\Omega (0)$
 increases when the star is in co-moving phase. This agrees with the crossover
 time estimate given in Eq.~(\ref{E:Weak_Field_Om}). However, if the star is in
 the forward creeping phase, the push received by the flux tubes due to pinned
 vortices saturates (and thus, independent of $\Omega$). So in this case, the
 value of $\Omega (0)$ has little effect on the magnetic evolution of the star.
 I have verified this in some of our runs.
\par
\begin{figure}
 \plotone{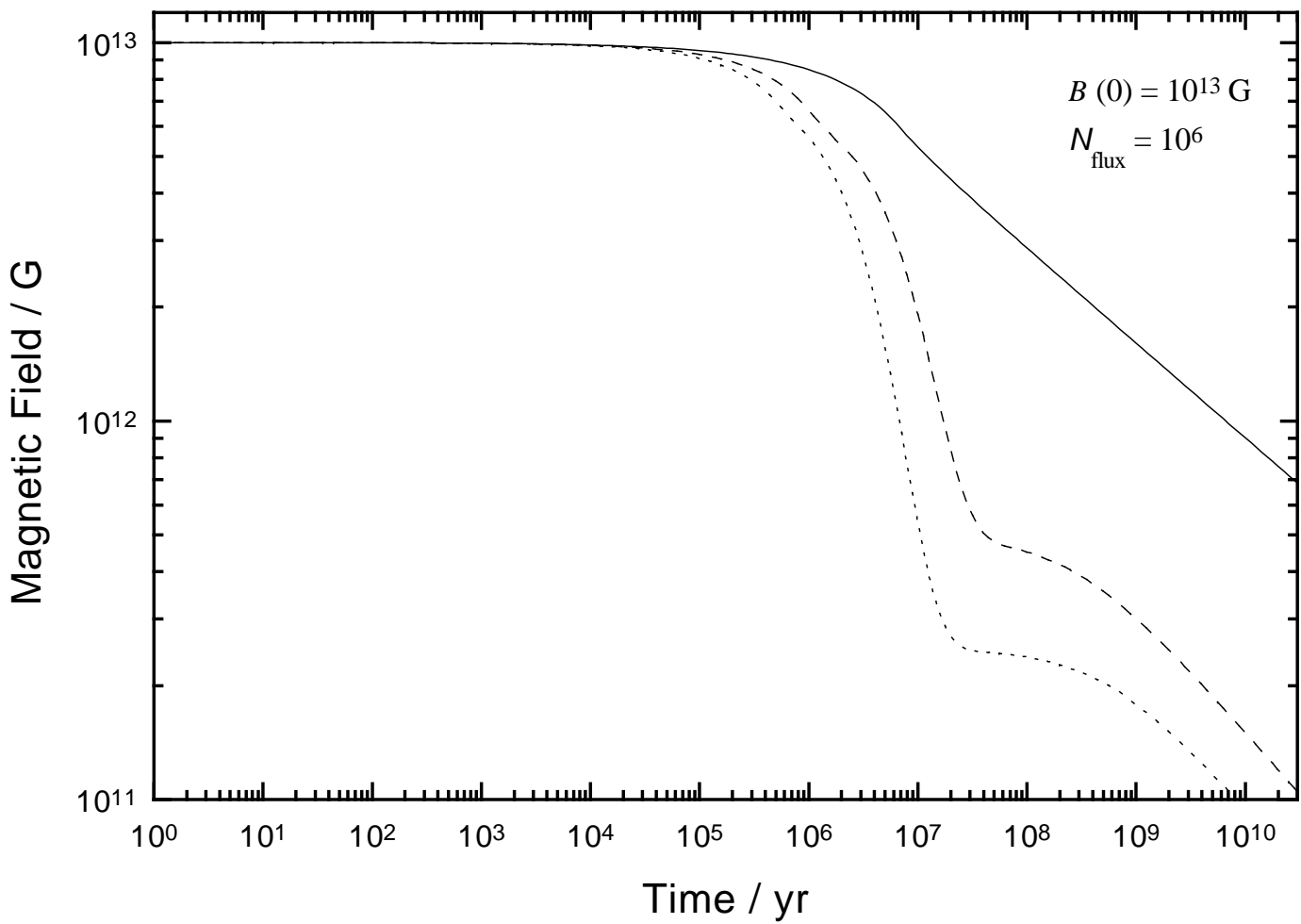}
 \caption{Magnetic field evolution as a function of the value $\gamma$. The
  solid, dotted and dash lines correspond to $\gamma =$ 0, 0.5, and 1
  respectively. Note that the initial magnetic field is set to $10^{13}$~G.
  \label{F:gamma}}
\end{figure}

 Next, we consider the effect of $\gamma$. Again, the only situation where
 the value of $\gamma$ can seriously affect the magnetic evolution of a star is
 when the pinning force is weak at some moment as compared to the buoyancy
 force; and this happens when both $B_{\rm star} (0)$ and ${\cal N}_{\rm flux}$
 are large. Fig.~\ref{F:gamma} show the magnetic field evolution when
 $B_{\rm star} (0) = 10^{13}$~G and ${\cal N}_{\rm flux} = 10^6$. We see that
 the evolution profile depends quite sensitively on $\gamma$. When $\gamma =
 0$, we observe the $B_{\rm star} \sim t^{-1/4}$ behavior at late times,
 indicating that the star is in the co-moving phase. This is reasonable since
 the only force that push the flux tubes out from the core comes from the
 vortex bundles. But when $\gamma \gtrsim 0.5$, buoyancy force becomes the
 dominant. Thus, the magnetic evolution follows an  exponential decay when $t
 \lesssim 10^6$~yr and $B_{\rm star} \sim t^{-1/3}$ at $t \gtrsim 10^9$~yr,
 indicating that the presence of forward creeping and reverse creeping phase at
 early and late times respectively.
\par
\begin{figure}
 \plotone{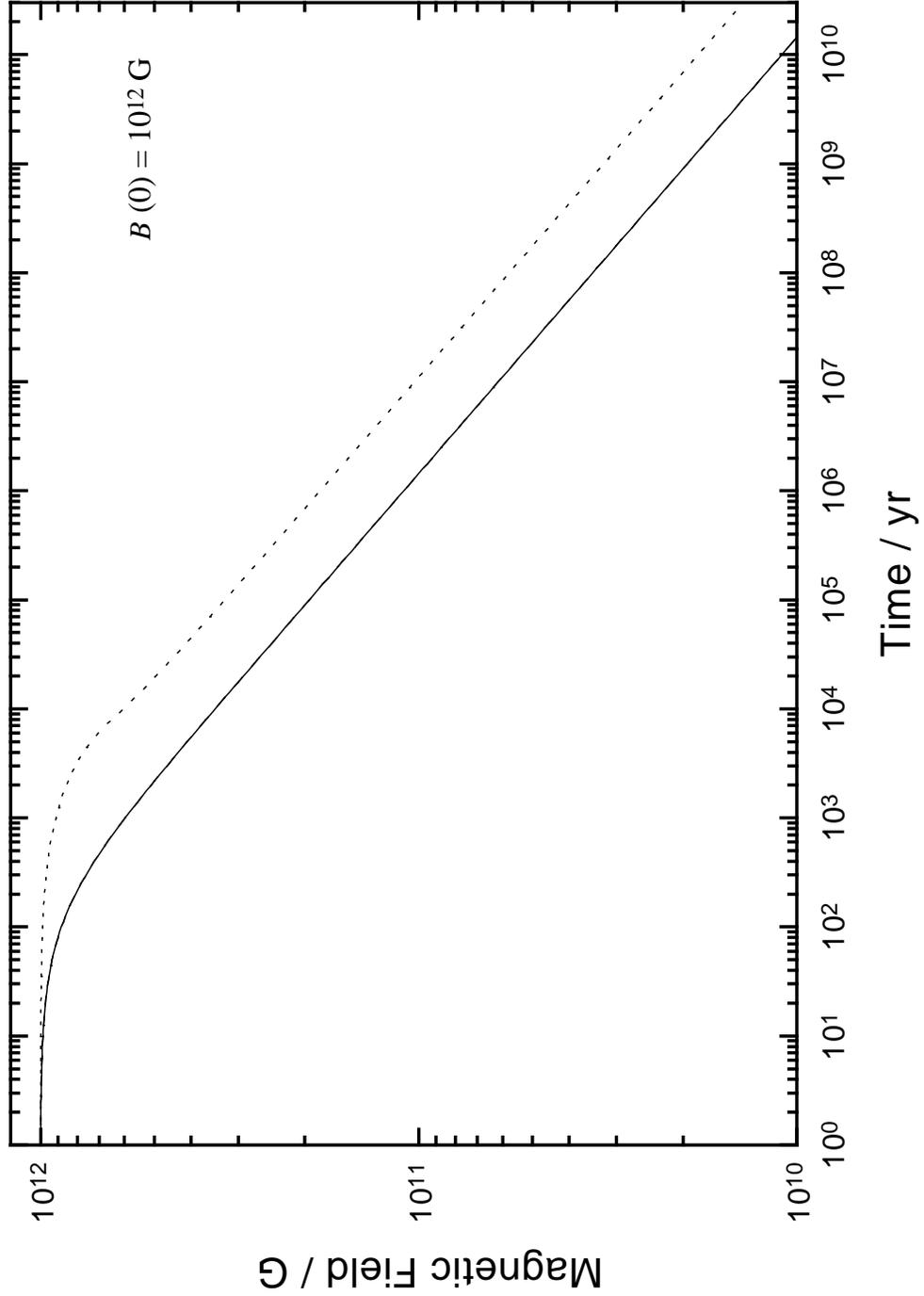}
 \caption{Magnetic field evolution as a function of ${\cal N}_{\rm flux}$. The
  solid, dash, and dotted lines correspond to ${\cal N}_{\rm flux} =$ 1, $10^3$
  and $10^6$ respectively. (a) shows the field decay when $B(0) = 10^{12}$~G,
  and (b) shows the field decay when $B(0) = 10^{13}$~G. Note that the solid
  curve overlaps with the dash one in (a). \label{F:flux}}
\end{figure}

 Finally, we consider the effect of ${\cal N}_{\rm flux}$. For $B_{\rm star}
 (0) = 10^{12}$~G, Fig.~\ref{F:flux}a tells us that the behavior of the star
 does not have any visible change ${\cal N}_{\rm flux}$ increases from 1 to
 $10^3$. They are all in the co-moving phase. This observation is consistent
 with the fact that $t_0$ in Eq.~(\ref{E:Weak_Field_Om}) is independent of
 ${\cal N}_{\rm flux}$. It is only when ${\cal N}_{\rm flux}$ increases to
 about $10^6$ we begin to see a slow down of field decay due to the presence of
 a relatively long period of forward creeping phase before the star eventually
 enters the co-moving phase.
\par
 The magnetic evolution is more dramatic if we take $10^{13}$~G as the initial
 field. Fig.~\ref{F:flux}b shows that when ${\cal N}_{\rm flux} = 1$, the star
 is basically in the co-moving phase all the time. When ${\cal N}_{\rm flux} =
 10^3$, a delay in field decay is observed due to presence of a long period of
 forward creeping phase at early times. Finally, when ${\cal N}_{\rm flux} =
 10^6$, the star switches to an initial exponential, followed by an eventual
 power law delay mode, indicating that the star has locked into the forward
 creeping $\longrightarrow$ co-moving $\longrightarrow$ reverse creeping
 pattern.
\par
 In summary, we find that in all reasonable parameter range, the value of the
 initial magnetic field will be reduced to $1/e$ of its original value in less
 than 1~Myr time when ${\cal N}_{\rm flux} \sim 1$ or when $B_{\rm star} (0)
 \lesssim 10^{12}$~G. If ${\cal N}_{\rm flux} \sim 10^6$, the characteristic
 field decay time is less than 20~Myr provided that $B_{\rm star} (0) \lesssim
 3\times 10^{12}$~G or $\gamma \sim 0$. The only way we obtain a characteristic
 decay time longer than 20~Myr is by using ${\cal N}_{\rm flux}$ as high as
 $10^6$, $B_{\rm star} (0)$ as high as $10^{13}$~G, and $\gamma \sim 0$ (see
 Fig.~\ref{F:gamma}) --- a set of physically unlikely combination of
 parameters.
\subsection{Observational Implications Of Strange Star Field Decay
 \label{S:Obs_Imp}}
 The presence of a large number of systematic and random errors in the
 observed pulsar sample together with the fact that kinematic age sometimes
 does not truly reflect the real age of a pulsar greatly complicate the
 analysis of pulsar magnetic field decay. Nonetheless, current statistical
 analysis and computer simulation of galactic pulsar distribution suggest that
 the characteristic field decay time is at least 20~Myr (\cite{Decay1};
 \cite{Decay2}; see also the recent review by Bhattacharya and Srinivasan
 1995).
\par
 Various authors have studied theoretically the possibility of a magnetic field
 decay in the core of a neutron star (\cite{Sauls_Review}; \cite{Srinivasan};
 \cite{Chau92}; \cite{Ding93}). However, Pethick and Sahrling (1995) pointed
 out recently that the field decay time in the inner crust of a neutron star is
 at least $80$~Myr due to the slow diffusion of the field through the inner
 crust. Consequently, core magnetic field decay only results in transporting
 the field from the stellar core to the inner crust; the total stellar field
 remains unchanged in $\gtrsim 80$~Myr. Therefore, the conventional hypothesis
 that pulsars are neutron stars is consistent with the observed field decay
 time of the star.
\par
 On the other hand, suggested by our numerical finding in \S\ref{S:Decay_Num},
 the magnetic field of a strange star decays with a characteristic times $\leq
 20$~Myr. In fact, Eq.~(\ref{E:tau_buoy}) tells us that the characteristic time
 for the buoyancy force dominated field decay $\tau_{\rm buoy}$ equals $4.2
 \times 10^{-3} {\cal N}_{\rm flux}$~Myr. Thus, $\tau_{\rm buoy}$ is longer
 than 20~Myr only when ${\cal N}_{\rm flux} \geq 2\times 10^7$ --- a value
 which is attainable only when the quark superconductor is extremely type~I.
 Therefore, our finding is inconsistent with the proposition that pulsars are
 strange stars.
\section{Conclusions \label{S:Conclude}}
 In summary, I consider the rotation of a multi-superconducting species
 object. The rotation of such an object is made possible by the formation of
 vortices similar to that of superfluid helium. This finding implies the
 existence of vortex bundles in the core of a rotating superconducting strange
 star. Because it is energetically favorable for the vortex bundles to pin to
 magnetic flux tubes, the rotational dynamics and the magnetic field evolution
 of a strange star are coupled.
\par
 Because the nuclear crust of a strange star cannot sustain a strong magnetic
 stress, the magnetic field evolution of the star is dictated by dynamics of
 the flux tubes in the core. The core magnetic field evolution due to
 inter-pinning of magnetic flux tubes and vortex bundles, and clumping of flux
 tubes is computed numerically. I find that in all reasonable parameter range,
 the characteristic decay time of the magnetic field is $\leq 20$~Myr. This
 finding does not agree with the hypothesis that pulsars are superconducting
 strange stars because current pulsar data strongly suggest that pulsar
 magnetic fields do not decay in 20~Myr.
\par
 A number of interesting questions remains. The phase diagram of interacting
 multiple component superconducting species in rotation remains unclear. It is
 also interesting to study the effects of alignment torque on magnetic and
 rotational history of a strange star, and the possible collapse of the nuclear
 matter crust due to magnetic stress. I plan to report them in my future works.
\acknowledgments
 I would like to thank Mal Ruderman for bringing up this problem to my
 attention and K. Y. Ding for providing us her field decay program for neutron 
 stars for reference. I would also like to thank H.-K. Lo, Geoff Ravenhall and
 Frank Wilczek for their useful discussions. The hospitality of Aspen Center
 for Physics is acknowledged where the early part of this work was performed in
 the summer of 1995. This work is supported by DOE grant DE-FG02-90ER40542.
\appendix
\section{Solution To The Strongly Interacting Superconducting Species Problem
 \label{S:App}}
 We prove the following claim by explicitly construct a solution to the
 problem. For simplicity, we shall drop all the star superscripts over the
 masses and charges.
\par\noindent
{\em Claim:} Suppose there is a finite number of superconducting species. Then
 Eq.~(\ref{E:Strong_Limit}) has a solution with $K \neq 0$ (and hence it has
 infinitely many solutions) if and only if there exist two vectors ${\bf v}_1
 \equiv (M_1, Q_1)$ and ${\bf v}_2 \equiv (M_2, Q_2)$ among the $(m_j, q_j)$
 together with two rational numbers $\alpha_1$, $\alpha_2$ such that $(m_j, q_j
 ) = \beta_{j1} {\bf v}_1 + \beta_{j2} {\bf v}_2$, with $\{ 1, \beta_{j1},
 \beta_{j2} \}$ being linear dependent over the set of all rational numbers
 $\Bbb Q$ for all $j$. In addition, $\alpha_1 \beta_{j1} + \alpha_2 \beta_{j2}$
 are rational numbers for all $j$, and $\alpha_1 Q_2 \neq \alpha_2 Q_1$.
\par\medskip\noindent
{\em Proof:} ($\Rightarrow$) By replacing vectors ${\bf v}_i$ by ${\bf v}_i /
 \ell$ ($i=1,2$) where $\ell$ is the least common multiple of the denominators
 of $\alpha_1 \beta_{j1} + \alpha_2 \beta_{j2}$, then $\beta_{ji}$ will be
 replaced by $\ell\beta_{ji}$. In addition, we can replace $\alpha_i$ by
 $\lambda\alpha_i$ where $\lambda$ is the least common multiple of the
 denominators of $\alpha_i$ and $\alpha_1 \beta_{j1} + \alpha_2 \beta_{j2}$.
 Then it is easy to verify that $\lambda\alpha_i$ and $\ell ( \alpha_1
 \beta_{j1} + \alpha_2 \beta_{j2} )$ are all integers. Thus, we can always
 assume that $\alpha_1$, $\alpha_2$ and $\alpha_1 \beta_{j1} + \alpha_2
 \beta_{j2}$ to be integers instead of rational numbers.
\par
 Suppose ${\bf v}_1$ and ${\bf v}_2$ are linear dependent on each other, then
 clearly the ratio of $m_j$ to $q_j$ are the same for all $j$. So, by choosing
 $N_j = 0$ for all $j$, $K = 1$, and $\Phi_v = m_j c / q_j$,
 Eq.~(\ref{E:Strong_Limit}) is satisfied. Now, we consider the more interesting
 case when ${\bf v}_1$ and ${\bf v}_2$ are linear independent vectors. Consider
 the equation
\begin{equation}
 \left( \begin{array}{c} P_1 \\ P_2 \end{array} \right) = \left(
 \begin{array}{cc} M_1 & -Q_1 \\ M_2 & -Q_2 \end{array} \right) \, \left(
 \begin{array}{c} K / h \\ \Phi_v / hc \end{array} \right) \mbox{~.}
 \label{E:A_1}
\end{equation}
 The linear independence of ${\bf v}_1$ and ${\bf v}_2$ implies that $M_1 Q_2 -
 M_2 Q_1 \neq 0$, and the solution of the above equation is given by
\begin{equation}
 \left( \begin{array}{c} K / h \\ \Phi_v / hc \end{array} \right) = \frac{1}{
 M_2 Q_1 - M_1 Q_2} \left( \begin{array}{c} Q_1 P_2 - Q_2 P_1 \\ M_1 P_2 - M_2
 P_1 \end{array} \right) \mbox{~.} \label{E:A_2}
\end{equation}
 Clearly, by choosing $P_1 = \alpha_1$ and $P_2 = \alpha_2$, we can check that
 the $K$ given by Eq.~(\ref{E:A_2}) is non-zero.
\par
 Now for each $j$, it is easy to check that by choosing $N_j = \alpha_1
 \beta_{j1} + \alpha_2 \beta_{j2}$ (which is an integer, and this is possible
 because $\{ 1, \beta_{j1}, \beta_{j2} \}$ is linear dependent over $\Bbb Q$)
 together with $K (\neq 0)$, $\Phi_v$ given in Eq.~(\ref{E:A_2}), then
 Eq.~(\ref{E:Strong_Limit}) is satisfied. (Clearly, if $N_j$, $K$, $\Phi_v$ is
 a solution of Eq.~(\ref{E:Strong_Limit}), then so is $\lambda N_j$, $\lambda
 K$, $\lambda \Phi_v$ for any non-zero integer $\lambda$. Thus,
 Eq.~(\ref{E:Strong_Limit}) has either no solution, or infinitely many
 solutions.) \hfill $\Box$
\par
 ($\Leftarrow$) On the contrary, suppose that for any linear independent
 vectors ${\bf v}_1$ and ${\bf v}_2$ chosen among $(m_j, q_j)$, and any
 rational numbers $\alpha_1$ and $\alpha_2$, we can find an $j$ such that $(m_j
 , q_j) = \beta_{j1} {\bf v}_1 + \beta_{j2} {\bf v}_2$ with either (1) $\{ 1,
 \beta_{j1}, \beta_{j2} \}$ being linear independent over $\Bbb Q$; or (2)
 $\alpha_1 \beta_{j1} + \alpha_2 \beta_{j2}$ is irrational; or (3) $\alpha_1
 Q_2 = \alpha_2 Q_1$. Now we analyze these three cases one by one:
\par
 Case~(1): If $\{ 1, \beta_{j1}, \beta_{j2} \}$ is linear independent over
 $\Bbb Q$, then from Eq.~(\ref{E:A_1}), we know that $N_j = P_1 \beta_{j1} +
 P_2 \beta_{j2}$. In order that $P_1, P_2$ and $N_j$ are all integers, the only
 possibility is $P_1 = P_2 = N_j = 0$. However, from Eq.~(\ref{E:A_2}), this
 implies $K = 0$ and hence Eq.~(\ref{E:Strong_Limit}) has no solution for $K
 \neq 0$.
\par
 Case~(2): If $\alpha_1 \beta_{j1} + \alpha_2 \beta_{j2}$ is irrational, then
 similar to the argument in case~(1), $N_j = \alpha_1 \beta_{j1} + \alpha_2
 \beta_{j2}$ is not an integer. Hence, this choice of $\alpha_1$ and $\alpha_2$
 does not produce a solution for Eq.~(\ref{E:Strong_Limit}).
\par
 Case~(3): If $\alpha_1 Q_2 = \alpha_2 Q_1$, then $K = 0$ from
 Eq.~(\ref{E:A_2}), which is impossible. \hfill $\Box$
\par \bigskip
 The above claim implies that there is, in general, no solution to the strongly
 interacting limit when the number of different species of superconducting
 Cooper pairs is greater than or equal to three. On the other hand, solution of
 Eq.~(\ref{E:Strong_Limit}) always exists when there are only two species of
 Cooper pairs, as in the case of strange matter.

\newpage
\begin{figure}
 \plotone{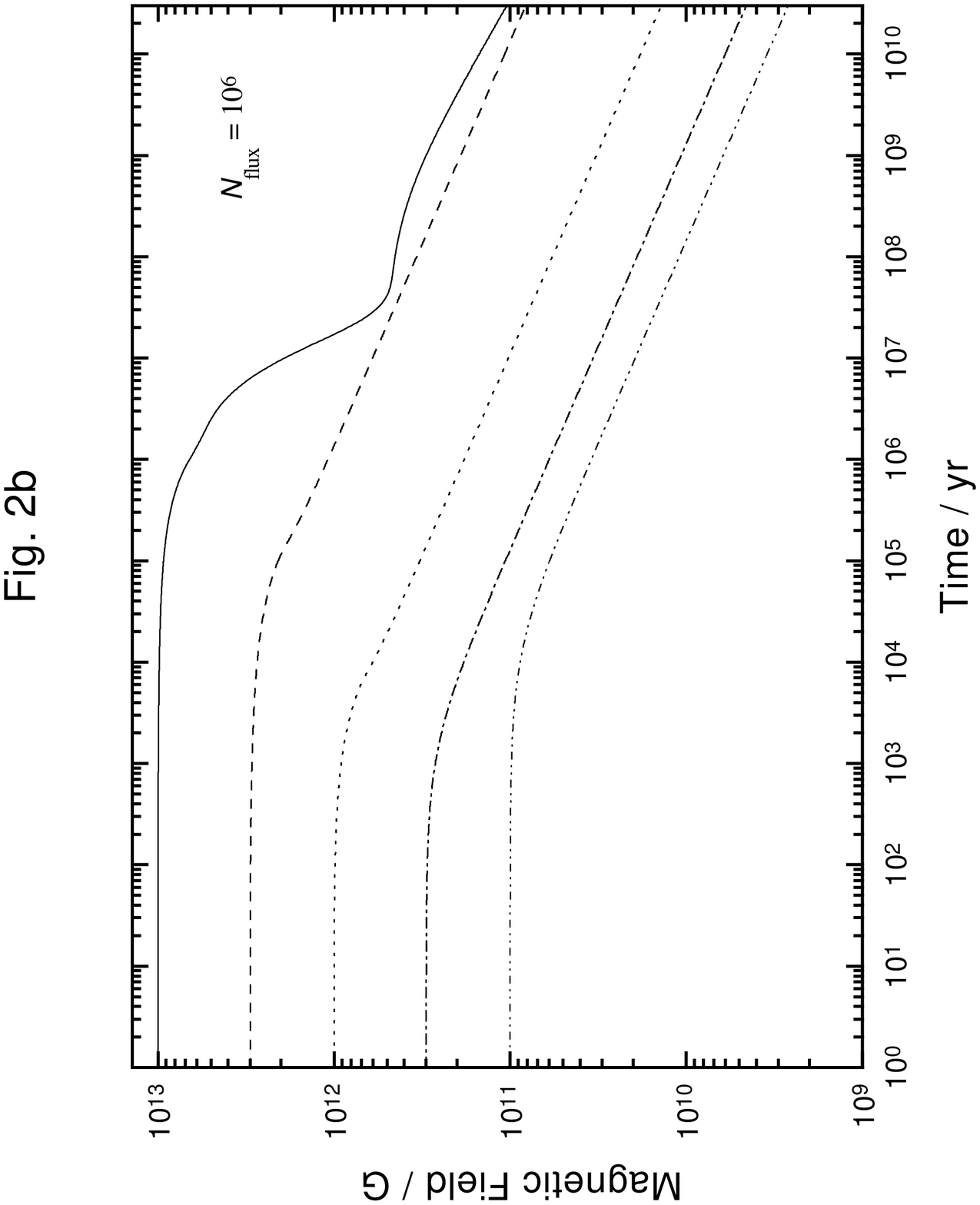}
\end{figure}
\begin{figure}
 \plotone{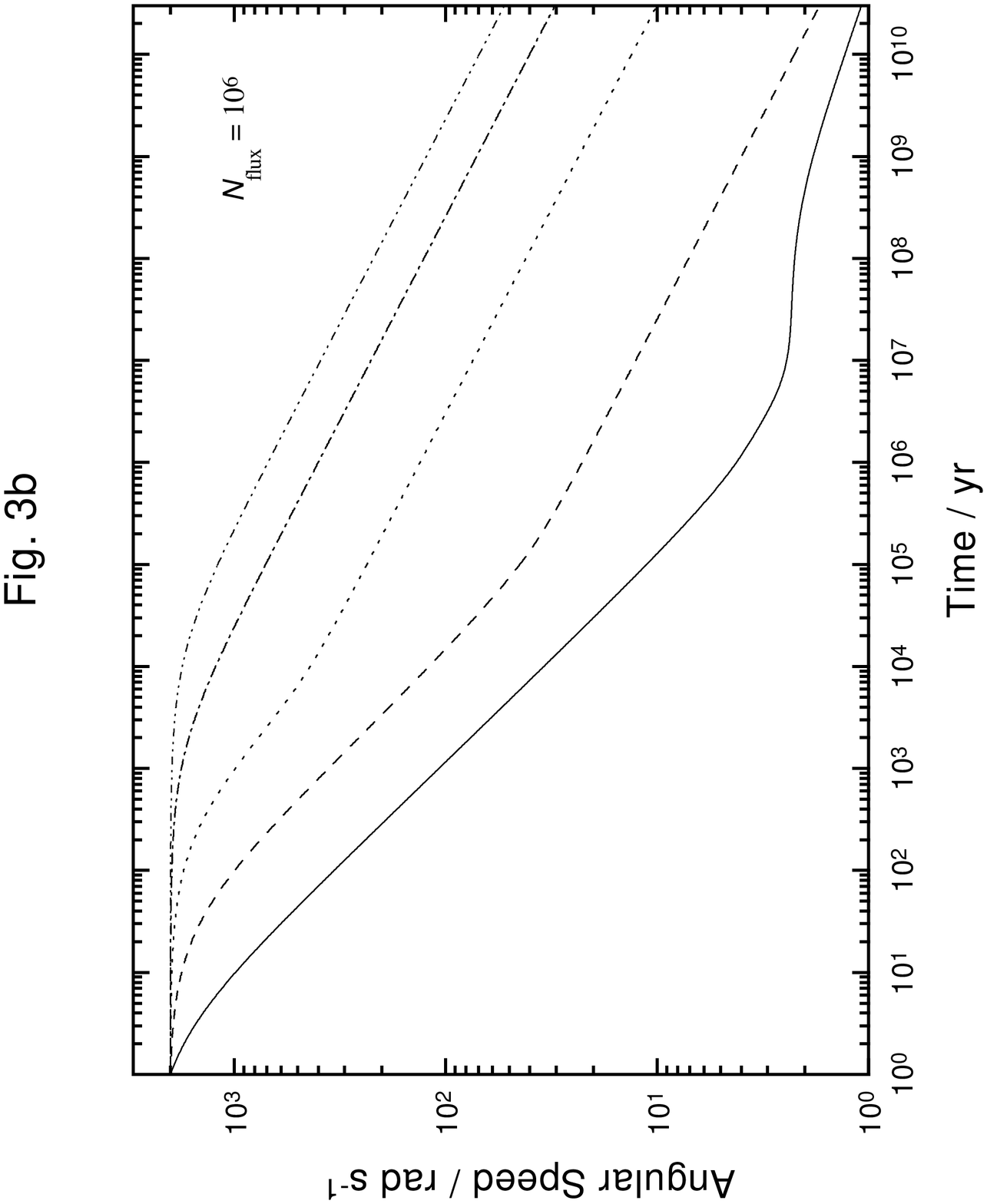}
\end{figure}
\begin{figure}
 \plotone{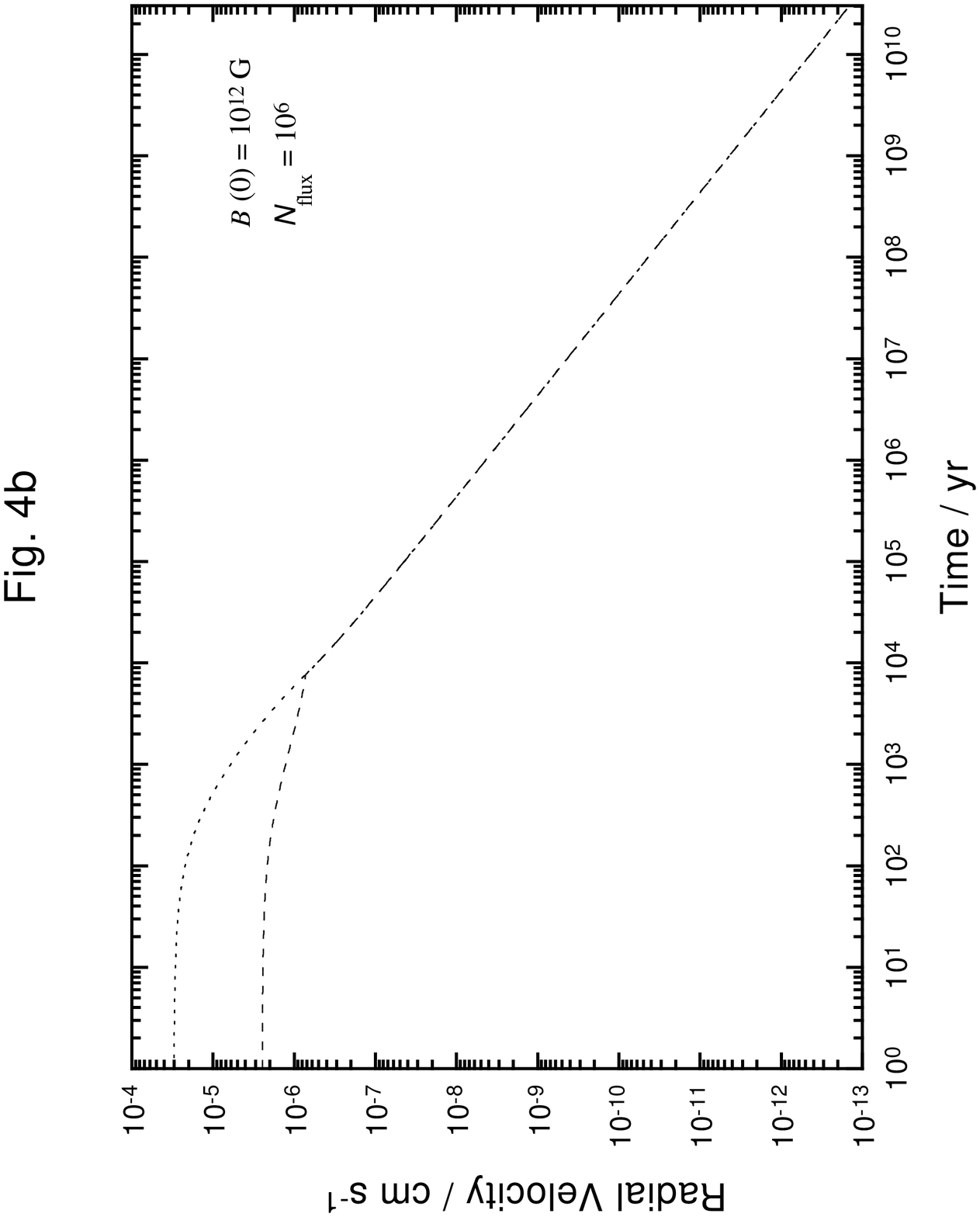}
\end{figure}
\begin{figure}
 \plotone{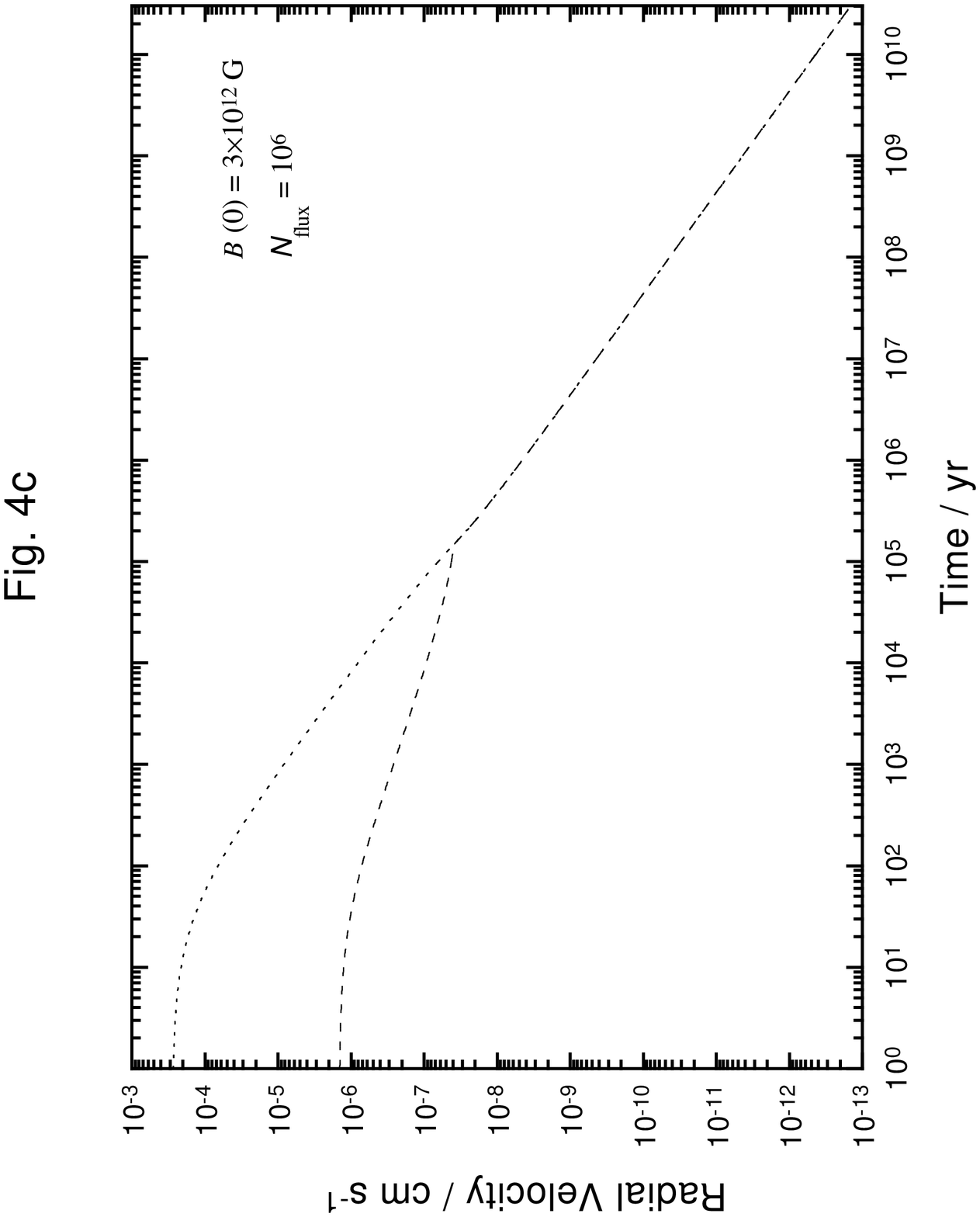}
\end{figure}
\begin{figure}
 \plotone{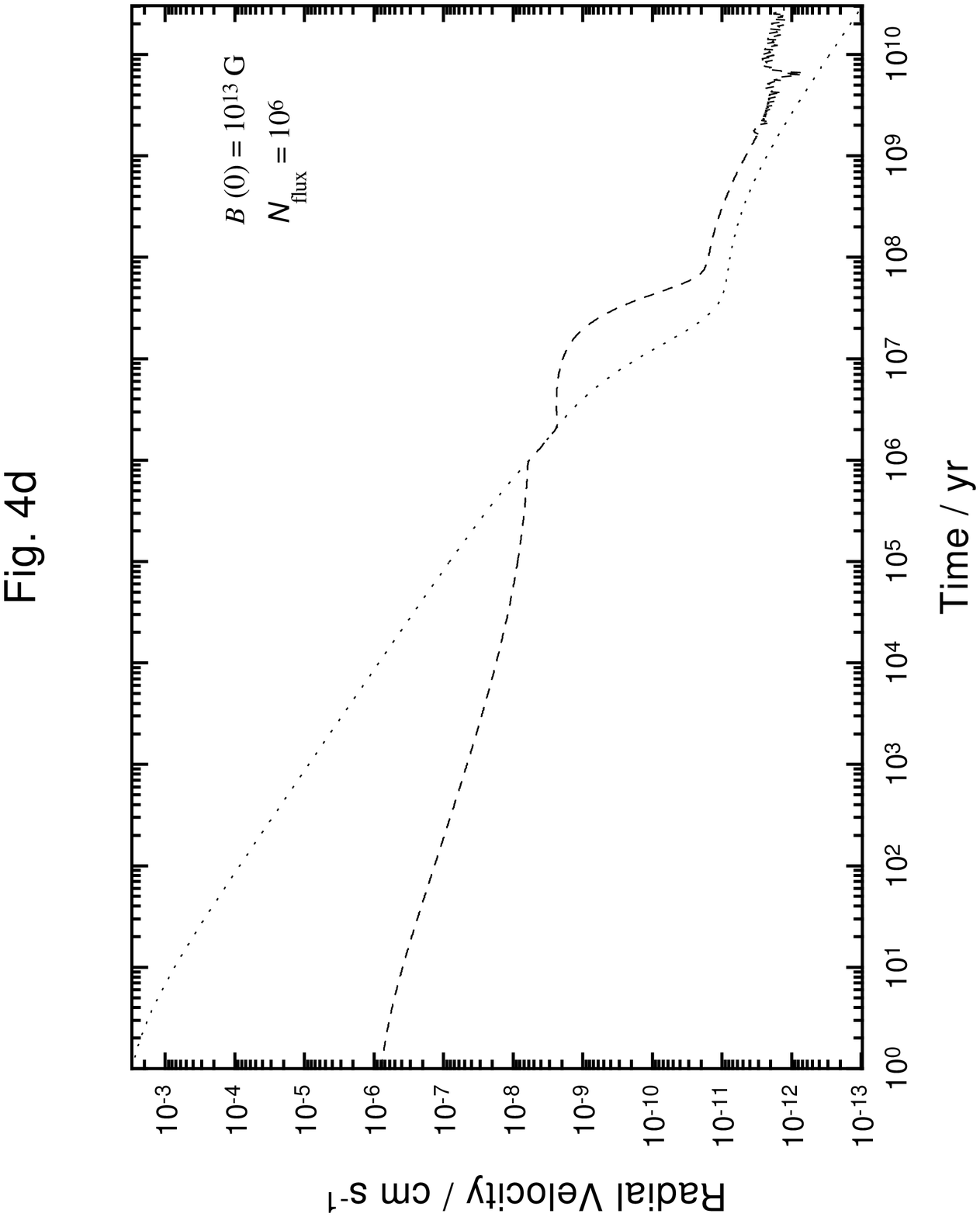}
\end{figure}
\begin{figure}
 \plotone{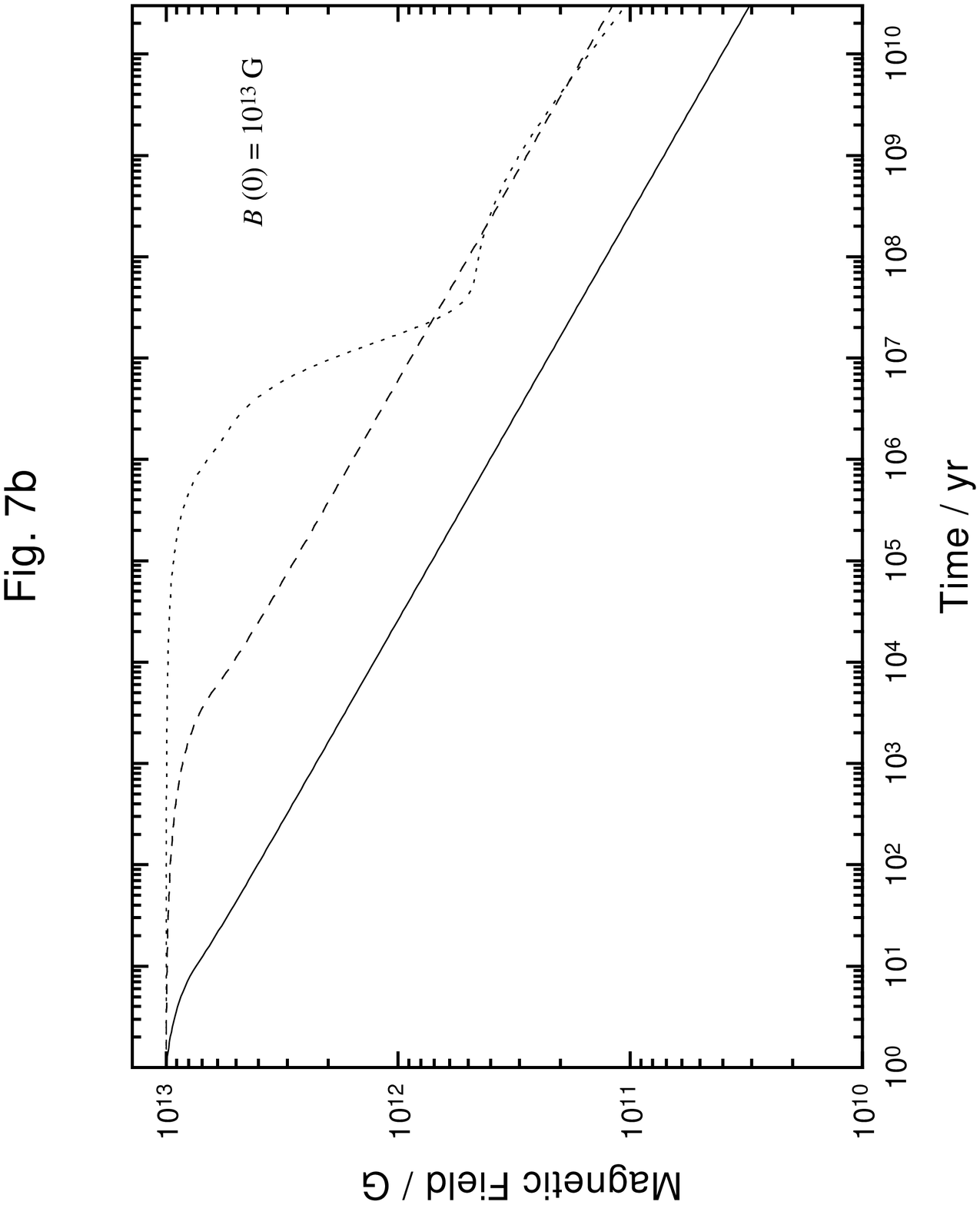}
\end{figure}
\end{document}